\documentclass[10pt,journal,compsoc]{IEEEtran}
\usepackage{amsmath,amsfonts}
\usepackage{algorithmic}
\usepackage{algorithm}
\usepackage{array}
\usepackage[caption=false,font=normalsize,labelfont=sf,textfont=sf]{subfig}
\usepackage{textcomp}
\usepackage{stfloats}
\usepackage{url}
\usepackage{verbatim}
\usepackage{graphicx}
\usepackage{cite}
\usepackage{multirow}
\usepackage[table,xcdraw]{xcolor}
\usepackage{tabularx}
\newcommand{\icon}[1]{\includegraphics[height=12pt]{#1}}
\newcommand{\changed}[1]{#1}

\usepackage{tikz}

\newcommand\submittedtext{%
  \footnotesize This work has been submitted to the IEEE for possible publication. Copyright may be transferred without notice, after which this version may no longer be accessible.}

\newcommand\submittednotice{%
\begin{tikzpicture}[remember picture,overlay]
\node[anchor=south,yshift=10pt] at (current page.south) {\fbox{\parbox{\dimexpr0.65\textwidth-\fboxsep-\fboxrule\relax}{\submittedtext}}};
\end{tikzpicture}%
}

\begin{document}
\title{Discrete Virtual Rotation in Pointing vs. Leaning-Directed Steering Interfaces: A Uni vs. Bimanual Perspective}

\author{Daniel Zielasko, Maximilian Späth, and \changed{Matthias Wölwer}

 \IEEEcompsocitemizethanks{\IEEEcompsocthanksitem D. Zielasko, M. Späth, and M. Wölwer are with Trier University, Germany. \protect\\
    E-mail: zielasko@uni-trier.de}
}



\IEEEtitleabstractindextext{
\begin{abstract}
In this work, we explore the integration of discontinuous Orientation Selection into steering interfaces intending to preserve the seamless sensation of real-world movement, while mitigating the risk of inducing cybersickness. 
Our implementation encounters conflicts in standard input mappings, prompting us to adopt bimanual interaction as a solution.
Recognizing the complexity that may arise from this step, we also develop unimanual alternatives, e.g., utilizing a Human-Joystick, commonly referred to as Leaning interface.
The outcomes of an empirical study centered around a primed search task yield unexpected findings. 
\changed{We observed a sample of users spanning multiple levels of gaming experience and a balanced gender distribution exhibit no significant difficulties with the bimanual, asymmetric interfaces. 
Remarkably, the performance of Orientation Selection is, as in prior work, at least on par with Snap Rotation.}
Moreover, through a subsequent exploratory analysis, we uncover indications that Pointing-Directed Steering outperforms embodied interfaces in usability and task load in the given setting. 
\end{abstract}

}

\maketitle
\IEEEpeerreviewmaketitle

\submittednotice

\IEEEraisesectionheading{\section{Introduction}\label{sec:introduction}}
Step-wise (discrete) virtual movement through virtual worlds often is a two-fold compromise. 
The first compromise is virtuality: there is often simply no space to physically move.
The second is that discrete movements seem more unnatural than continuous virtual movement, such as Pointing-Directed Steering, and have the potential to adversely affect the experience or break presence (plausibility) \cite{Bowman1997travel, Riecke2021teleportProCon, Bakker2003}.
Still, discrete virtual movements, for instance, in the form of teleportation \cite{Prithul2021teleportation}, are applied because of convenience or to prevent a perceptual conflict between the visual and the vestibular system, known to be a major contributor to cybersickness~\cite{LaViola2000, Buttussi2023}.
In most of these interfaces, only virtual translation is considered, which makes sense as physical rotation requires less space and is less often limited than physical translation.
However, there are a lot of situations where physical rotation is also restricted, for example, with the user sitting at a desk~\cite{Zielasko2019deskTravel}, on a couch, or in transit~\cite{Schmelter2020}.
There is research indicating that the integration of discrete virtual rotation with continuous translation \changed{(steering)} looks promising~\cite{Sargunam2018, Arns2004}.
Furthermore, Zielasko et al.~\cite{Zielasko2022rotation} recently found a pointing-based rotation approach working better--excluding spatial orientation--than classical direction-based ones in a stationary rotation task.
In their study, it remains unclear whether this approach can be seamlessly integrated into a fully virtual travel interface encompassing both translation and rotation.
Therefore, we aim to delve deeper into the combination of discrete rotation with continuous translation with the study that is presented here.
This leads us to our first research question:

\textbf{RQ1:} \textbf{How do different forms of discrete virtual rotation integrate with steering interfaces, specifically with regard to performance, usability, spatial orientation, and task load?}

\changed{This is of particular interest as steering interfaces, traditionally translate the user in a continuous form}.
We integrated direction-based and pointing-based rotation into \changed{traditional} steering interfaces \changed{(translation)} to answer this question.
One obstacle that occurs when combining existing implementations of stationary rotation~\cite{Farmani2018snapping, Zielasko2022rotation} with steering interfaces is an overload of physical inputs, and thus potentially cognitive resources.
For instance, the consolidation of pointing-based rotation and pointing-based translation can lead to a conflict of (optimal) controller/input mappings that can be solved in different ways.
This introduces our second research question:

\textbf{RQ2:} \textbf{Which rotational and translational input mappings work the best for Pointing-Directed Steering and specifically require the least amount of mental demand?}

We implemented different solutions, including the use of two controllers, a combination of thumbstick and device-based pointing, and a Human-Joystick translation interface, to answer this question.
For translation, Human-joystick or Leaning interfaces show high potential in a number of preliminary works in key characteristics like kinesthetic, proprioceptive, and vestibular stimulation \cite{Guy15, WangLindemann2012, ValkovWIM2010, Marchal2011Joyman}, presence \cite{Freiberg2017, Marchal2011Joyman}, task performance \cite{Mielhbradt2018, Harris2014, Nguyen19}, and engagement \cite{Zielasko2020LookAround, KitsonRiecke2015, Marchal2011Joyman}.
Due to their characteristical \textit{embodied} or body-centered/based interaction, these interfaces additionally are considered intuitive and natural. 
This makes them perfect candidates to solve the above-mentioned mapping conflicts.
Moreover, these approaches are seldom assessed against pointing-based interfaces—a comparison we aim to address in this paper.

In the remainder of this work, we introduce and discuss existing work in the field of discontinuous translations and rotations as well as steering interfaces more broadly (see Section~\ref{sec_rw}). 
Afterward, we present our scientific methodology to answer the proposed questions in Section~\ref{sec_method}.
This includes hypotheses, a detailed description of the interface implementations, and the experimental design of the conducted user study.
The results are presented in Section~\ref{sec_results} and finally discussed in Section~\ref{sec_discussion}.

\section{Background \& Related Work\label{sec_rw}}
We introduce steering interfaces in Section~\ref{sec_rw_steering}. 
Then there is a discussion of research around discrete rotation (see Section~\ref{sec_rw_rotation}).

\subsection{Steering\label{sec_rw_steering}}
Next to \textit{natural} travel interfaces, such as real walking, redirected walking, or walking-in-place, steering is, together with teleportation, one of the most common virtual travel paradigms \cite{laViola2017_3DUI}.
Teleportation is usually considered to describe discrete virtual travel.
This includes selection/pointing-based target specification \cite{Bozgeyikli2016} and less common, direction-based target specification.
Steering involves continuous virtual travel, and there are also hybrid methods documented in prior studies \cite{Adhikari2021lean, Bhandari2018dash}.
In the context mentioned above, steering represents a compromise when the aspiration for natural virtual locomotion surpasses the physically achievable limits.

Steering is typically characterized by the decomposition into 1) a direction component $dir$ (usually $\in \mathbb{R}^2$ for ground-based or $\in \mathbb{R}^3$ for \textit{flying} interfaces), 2) a speed/power/motor component $speed \in \mathbb{R}$, and 3) the input component \cite{Bowman1997travel, Brument2021}.
To this list we want to add 4) the transfer functions that map the latter to direction and speed. 
According to the definition by La Viola et al. \cite{laViola2017_3DUI}, in a steering interface the directional component is under the continuous control of the user (direction-specification) but for most implementations, this is also true for the motor component (speed-specification).

In practice, $speed$ is predominantly determined by applying a transfer function to a directly controlled input parameter of a one-dimensional axis input, i.e., $\in [-1,1]$ or $\in [0,1]$.
This input can also be virtual, as observed in Human-Joystick interfaces (see Section~\ref{sec_rw_leaning}).
In the easiest cases, the transfer function has linear behavior or represents a step function over constants (e.g., \{zero speed, maximum speed\}).
More complex cases are quadratic \cite{vonKapri2011}, power functions \cite{Flemming2022, Adhikari2022hyperjump} or non-continuous functions, for example when deadzones are introduced \cite{Zielasko2016HMDnav, Zielasko2020LookAround}.
Gao et al. discuss the influence of different transfer functions in embodied locomotion interfaces \cite{Gao2021}.
Rarer, but possible are also acceleration-based interfaces and/or interfaces that are not interpolated between a minimum and maximum speed.

\subsubsection{DualStick (Gamepad-like)\label{sec_rw_dual}}
Desktop-based 3D video games, when dealing with a rotationally stationary user, are traditionally steered using two distinct 2D input axes—one for translation and one for rotation (DualStick steering).
Both axes define direction and speed simultaneously.
These axes are controlled by, e.g., thumbsticks, joysticks, track-pads, cross-pads, or keyboard (WASD) \& mouse.
Some of them generate continuous/analog input and others binary/discrete/digital input.
In the case of binary inputs, the speed component is more complex to manipulate:
Sometimes there are also only two target states: full speed and standstill.
These can either be directly mapped from the input or be switched between them. 
Another common solution for binary inputs is that the user does not directly control the speed and instead specifies acceleration. 
Then, through pulse modulation between 0 and maximum acceleration, with a little practice, different speeds can be achieved and maintained.
These DualStick interfaces are also used and evaluated in VR applications preferably but not necessarily if the user is stationary in rotation and translation \cite{Zielasko2016HMDnav, Zielasko2019deskTravel, Zielasko2020LookAround, Zhixin2016fingerWIP, Nabiyouni2015}.
The optimal use case for such interfaces is in ground-based movement, primarily owing to their restricted degrees of freedom (2 translations, 2 rotations). In the case of rotation, usually, only one axis is controlled (Yaw rotation), and extensions can be made—usually to Pitch—enabling basic flight, albeit with limitations.
At the very least, when one wishes to directly manipulate the "altitude" (relative to an imaginary horizon), it requires a minimum of 5 degrees of freedom, which necessitates incorporating additional inputs~\cite{Ortega2020Gamepad} (cf. control interfaces for drones, etc.), the mapping of translation or rotation across multiple sticks~\cite{Hashemian2020HeadJoystick}, or the separation of the direction specification, for instance, to an embodied direction component~\cite{Hashemian2020HeadJoystick} (see Section~\ref{sec_rw_embodied_dir}).
Finally, interfaces that require all 6 degrees of freedom, i.e., including Roll, are quite rare.

\subsubsection{Embodied Direction\label{sec_rw_embodied_dir}}
In the absence of virtual rotation, steering interfaces where the inputs for the translation's speed and direction are decoupled, are more frequent as the opportunity to embed the direction specification into the physical environment is used.
One example is Pointing/Pointer-Directed Steering \cite{Suma2007, Mine1995virtual, Bowman1997travel, vonKapri2011}.
In these interfaces, the movement direction is specified by the pose of a hand or tracked device, such as a wand or flying joystick, and the speed is often given by a 1D axis input, such as a trigger.
Certainly, the direction component can be replaced by other modalities that are considered, at times, more natural, such as the direction of gaze, the user's head orientation \cite{Suma2007, Ruddle2013, Zielasko2016HMDnav, Zielasko2019deskTravel}, the torso's orientation \cite{Zielasko2020LookAround}, chair orientation \cite{Adhikari2021lean, Hashemian2023}, and so forth.
This holds also for the speed component, which can be taken by body tilt \cite{Marchal2011Joyman}, hand gesture/position \cite{Mine1995virtual, Jeong2009}, the heel position \cite{Zielasko2016HMDnav}, arm swinging \cite{Wilson2016armswing}, or in-place step frequency.
The latter two are traditionally not classified as steering interfaces but rather as natural travel interfaces. 
However, this characterization hinges on the underlying features, and both inputs do not inherently contradict the definition of a steering interface, as outlined above.
All the mentioned components can be further combined, for instance, to complement or adapt interfaces for specific applications. 
The design space of possible implementations expands significantly when considering different degrees of freedom for translation and rotation, whether they are virtual or physical.

Finally, even different input modalities can be combined to form a single new one, so a common implementation of Pointing-Directed Steering with a flying joystick/wand/tracked hand device utilizes the device's pose but only indirectly.
The actual direction of travel is determined by a thumbstick/analog stick/kuli head's inclination/tilt, which is mounted on the device and is often oriented with respect to the underlying controller's pose, rather than being locally evaluated \cite{Freitag2016}.
In this configuration, the interface offers some advantages: it can be used flexibly, i.e., it behaves exactly as a regular Pointing-Directed Steering interface, when the analog stick is only used in its positive forward axis. 
For steering backward (or sideways, strafing), the user does not need to physically point to their back; instead, they can simply flex the analog stick backward (to the sides).
Nevertheless, we posit that this flexibility comes at the cost of increased complexity and a steeper learning curve for users. 
This complexity may limit the advantages of improved spatialization when compared to the straightforward approach of forced pointing with the whole arm.

\subsubsection{Human-Joystick (Leaning)\label{sec_rw_leaning}}
Another class of steering interfaces combines the aforementioned two and interprets the user's body as a single analog stick (2D axis input).
Human-Joystick interfaces, often also referred to as Leaning, take various forms in manipulating speed and/or travel direction based on the inclination of the user's body, either as a whole \cite{Marchal2011Joyman} or in parts \cite{Guy15, Zielasko2016HMDnav, Adhikari2022hyperjump}.
At this point,  it is important to note that in most implementations, the tilt of the body is not directly measured, as demonstrated in this work on torso-directed steering \changed{by Zielasko et al.} \cite{Zielasko2020LookAround}.
Instead, it typically involves an offset vector between the reference and current position of the user, as demonstrated in \cite{Zielasko2016HMDnav, Adhikari2022hyperjump}, or through a device that is moved by the user's body \cite{Kitson2017}. 
This makes a difference because it theoretically allows the interface to be used differently than intended by the metaphor, cf. interfaces that are more like stepping \cite{Nguyen19}, here referred to as Pivoting--based on the movement in basketball--or Balancing \cite{Kitson2017} than leaning/tilting. 
Furthermore, the input can behave non-linearly depending on the tracked body part. 
For example, the head moves on a circular path when the user bends at the hip, which is relevant in cases where the head is tracked.

Human-Joystick interfaces, demonstrated across a diverse range of implementations, consistently exhibit positive characteristics. One notable advantage is that users can keep their hands free for other interactions \cite{Zielasko2016HMDnav, Hashemian2021IsWN, Hashemian2023}. 
This attribute is especially crucial in the current context, as we aim to integrate solutions with proven advantages, recognizing that combining them may introduce potential conflicts in the input mapping.

Furthermore, these interfaces have been shown to generate kinesthetic, proprioceptive, and vestibular stimulation \cite{Guy15, WangLindemann2012, ValkovWIM2010, Marchal2011Joyman}, which, among others, have a positive impact on the perception of egocentric self-motion perception \cite{Kitson2017, Kruijff16, Riecke2006}.
The latter might also serve as an explanation for an observed increased sense of presence \cite{Freiberg2017, Marchal2011Joyman}, enjoyment, and engagement \cite{Zielasko2016HMDnav, Marchal2011Joyman, KitsonRiecke2015}.
Finally, Human-Joystick interfaces have demonstrated the ability to compete with or even surpass controller-based steering interfaces in terms of both performance and user experience metrics on multiple occasions~\cite{Harris2014, Mielhbradt2018, Nguyen19, Zielasko2016HMDnav}.

A notable weakness when interpreting the human body as an analog stick is that, in the absence of physical rotation, this approach alone does not constitute a complete steering interface and necessitates additional input for virtual rotation.
Solutions like Scrolling, i.e., utilizing part of the freedom in the head's Yaw \cite{Zielasko2016HMDnav, Langbehn2019turn}, do exist but are restricting.
However, this makes this type of interface a receptive candidate for a virtual rotation methodology that we want to explore.

\subsubsection{Comparisons\label{sec_rw_comparisons}}
Various evaluations exist that compare steering methods with each other or with other paradigms.
For instance, von Kapri et al. \cite{vonKapri2011} in a flying coin collection task with physical rotation specification find no significant difference in presence and usability between three interfaces: Pointer-Directed Steering (tracked device), Pointing/Hand-Directed Steering, and a standing Leaning condition called PenguFly.  
The Pointing-Directed Steering interface showed the highest performance.
PenguFly reported higher sickness ratings, although it stood out as the most precise interface.

Zielasko et al. \cite{Zielasko2016HMDnav} find a Leaning and a heel-based speed-specification performing overall the best \changed{for} a sitting user (virtual rotation-specification via Scrolling \cite{Langbehn2019turn} and direction specification via head orientation) and a combined naive search and maneuvering task, against a DualStick and a head-shaking interface.

Nguyen et al. \cite{Nguyen19} evaluate a head-directed touchpad, a seated Balancing, a standing Pivoting, and a walking interface in a naive search task.
All interfaces make use of physical rotation-specification.
The workload is higher with the non-spatialized touchpad interface than with physical walking and Pivoting. 
The task completion time is significantly slower with the touchpad than with all other methods.

Buttussi et al. \cite{Buttussi2021} evaluate three interfaces again with physical rotation specification: a seated Leaning interface, touchpad-based steering, and teleport in an informed search task.
For the steering interfaces the direction of travel is specified together with the speed by the respective 2D axis input.
Teleportation leads to the highest performance and the least sickness; the Leaning interface performs better than the touchpad-based steering.

Hashemian et al. \cite{Hashemian2023LeaningAndInteraction} conducted a comparison between real walking, an analog stick-based Pointing-Directed Steering interface, and two human-joystick interfaces in a dual-task design. In this setup, participants engaged in both travel and simultaneous interaction with virtual objects. Summarizing the observed effects on various dependent variables, real walking exhibited the best performance, while the pointing-based method performed the least effectively.

In summary, Human-Joystick interfaces in varying implementations often show either overall positive or at least positive in single key metrics, especially compared to other continuous travel methods such as non-spatialized methods that make use of a single 2-axis input (analog stick or touchpad). 
However, direct comparisons to Pointing-Directed Steering methods, which are arguably simple to use and also show applicability, are sparse.
Therefore, we want to further investigate our research questions with respect to these two paradigms.

\subsection{Discrete Rotations\label{sec_rw_rotation}}
As discussed above, continuous virtual rotation can be necessary when physical body rotation is not possible or not preferred but it also generates a significant amount of optical flow, which is a primary cause of cybersickness \cite{LaViola2000}. 
To address this challenge for translational movements, discrete movements such as teleportation are often utilized. 
At the same time, rotational movements seem to have a greater effect on sickness than translational movements, as indicated by the results of driving simulator experiments by Kemeny et al. \cite{Kemeny2017} and long-term observations by Zielasko et al \cite{Zielasko2021sicknessSubject}. 

Sargunam et al. \cite{Sargunam2018} apply the concept of discrete movements to virtual rotations combined with continuous translations and compare three different analog-stick-based virtual rotation techniques in terms of their impact on spatial orientation and cybersickness. 
They find that Rotational Snapping in increments of 30 degrees comes with significantly lower cybersickness than the continuous baseline condition (30 degrees/second) and the condition that utilizes field of view modifications \cite{fernandes2016, zielasko2018dynamic, bolas2017dynamic}, although they find no difference in spatial orientation.

\changed{Benda et al. \cite{Benda2023} have implemented a pointing-based Orientation Selection that is then executed with a blacked-out screen until the user reaches a comfortable head position again (\textit{Resetting}).
Participants preferred more direct rotation methods.}

\changed{Snap Rotations describe a series of small discrete viewpoint rotations in reaction to a continuous directional input.}
Farmani and Teather \cite{Farmani2018snapping} use semi-continuous Rotational Snapping with a mouse's x-axis as input and find that a rotation speed threshold of 25 degrees per second and a jump size of 22.5 degrees results in a 40\% reduction in cybersickness levels compared to pure (continuous) Scrolling. 
These findings were replicated in a second study where Rotation Snapping was combined with translation \cite{Farmani2020evaluating}.

Recent research by Zielasko et al. \cite{Zielasko2022rotation} has extended these findings. They discovered that pointing-based Orientation Selection, involving a simple point-and-turn approach, performs remarkably well in both informed and naive rotational search tasks. 
This performance holds true when compared to Snapping methods, with notable improvements in task performance and usability.
In the early stages of spatial updating, inaccuracies were detected when participants selected their orientation. However, these discrepancies do not persist and do not influence the final results.
In summary, this naturally prompts the question of whether these results can be replicated within a complete locomotion interface and how (\textbf{RQ1} \& \textbf{RQ2}). 
Moreover, in their study, they observed no improvement in spatial orientation when upper-body Leaning was used for the selection instead of a tracked controller. However, there was a decrease in perceived usability.

For the sake of completeness, it should be mentioned that another obvious solution for avoiding circular vection (perception of ego-motion) when specifying user orientation is the use of exocentric methods (TeleTurn). 
In the immersive context, these can be found, for example, in teleportation methods that allow for simultaneous \cite{Bozgeyikli2016, Bimberg2021anchor, Funk2019, Wolf2021AugmentedTeleport} or asynchronous \cite{Wolf2021AugmentedTeleport} position and orientation specification.
Less common are exocentric steering methods \cite{Griffin2019, Lugrin2019}. 
However, in this work, we focus on egocentric methods.


\section{Method\label{sec_method}}
Combining the results from related work, without expecting interaction effects, a discrete selection-based rotation combined with continuous Pointing-Directed Steering \changed{(translation)} should yield excellent results. 
However, it is often not that straightforward, and particularly in this case, we anticipate interactions leading to problems in interface usability due to increased complexity:
Naively, for both motion components (translation and rotation), a direction needs to be specified simultaneously and independently, which raises the need to operate two pointing devices at once. 
To address this potential issue, we examine additional degrees of freedom in the interface design, following the classification of Zielasko et al.~\cite{Zielasko2022rotation}, with a focus on reducing the aforementioned complexity. 
Consequently, we reintroduce Snap Rotation, even though it shows poorer performance when considered in isolation compared to selection-based rotation~\cite{Zielasko2022rotation}. 
In the realm of translation, we also explore Leaning interfaces in addition to Pointing-Directed Steering as they provide an additional source of 2D input that is hands-free. 
While Leaning interfaces were recently deemed less usable for rotation specification~\cite{Zielasko2022rotation}, this does not necessarily apply to translation~\cite{Zielasko2016HMDnav, Nguyen19, Hashemian2023}. 
Furthermore, there is a lack of direct comparisons between Pointing-Directed Steering and Leaning interfaces in the literature.

In the following, we implement and empirically examine the different combinations.
From a theoretical point of view, the aforementioned considerations lead to a study design of two factors with two levels each: translation (Pointing-Directed Steering vs. Leaning) and rotation-specification (discrete Orientation Selection vs. Snapping Turning).
The degrees of freedom in the specific implementation result in the branching of specific conditions we examine. 

\subsection{Interfaces\label{interfaces}}
First, we describe the different implementation components in Section~\ref{sec_implementation_pointing}-\ref{sec_implementation_snap} and then how they are composed to the different conditions in Section~\ref{sec_implementation_conditions}.
\changed{All discrete scene rotations were performed as a hard cut, without any form of transition effect, such as fade-to-back (cf. \cite{feld2024transition, woelwer2024posterFadeaway}). The main reason for this was that in the combination of continuous movement (translation) and discontinuous rotation, it would have massively disrupted the user experience to have limited visibility for a certain period during a rotation.}

\subsubsection{Translation: Pointing-directed Steering\label{sec_implementation_pointing}}
With Pointing-Directed Steering, the user's viewpoint moves in the direction of a tracked handheld controller's pose $\in \mathbb{R}^3$ (in global coordinates) that is projected to the ground plane.
The speed of the movement is always positive (or $0$) and determined by the primary index trigger.
Therefore, the trigger's input $\in [0,1]$ is linearly interpolated to $[0, 3]~m/s$. 

\subsubsection{Translation: Leaning\label{sec_implementation_leaning}}
For Leaning, a reference position of the user's head (HMD position) is calibrated $P_{ref}$.
\changed{To determine this reference position, the participants were briefly asked to sit relaxed and look forward before each Leaning condition. 
The experimenter then triggered the recording of the reference point. 
As such, we use, as Flemming et al. did \cite{Flemming2022}, the head's center of yaw rotation, which is a more reliable null position than the actual HMD position. 
During calibration, this is assumed to be $10~cm$ behind the HMD.}

In every frame, the offset $O \in \mathbb{R}^3$ between the current head position $P_{cur}$ and the reference position is determined $O = P_{ref} - P_{cur}$ and projected to the ground plane.
With Leaning, the user's viewpoint moves in this direction.
For the speed of the movement, the length of the offset $O \in \mathbb{R}^3$ is normalized and clamped at $0.2~m$. 
This scalar $x \in [0,1]$ is, with the use of a power function, interpolated between $[0, 3]~m/s$ using $x^{1.53}$.
Thus the user reaches the maximum speed when reaching or exceeding a distance of $0.2~m$ to the reference point.
This scaling in the past has been shown to successfully replace the use of deadzones in embodied interfaces \cite{Flemming2022, Adhikari2022hyperjump}.

\subsubsection{Rotation: Selection\label{sec_implementation_selection}}
\begin{figure}
  \centering
  \includegraphics[width=\columnwidth]{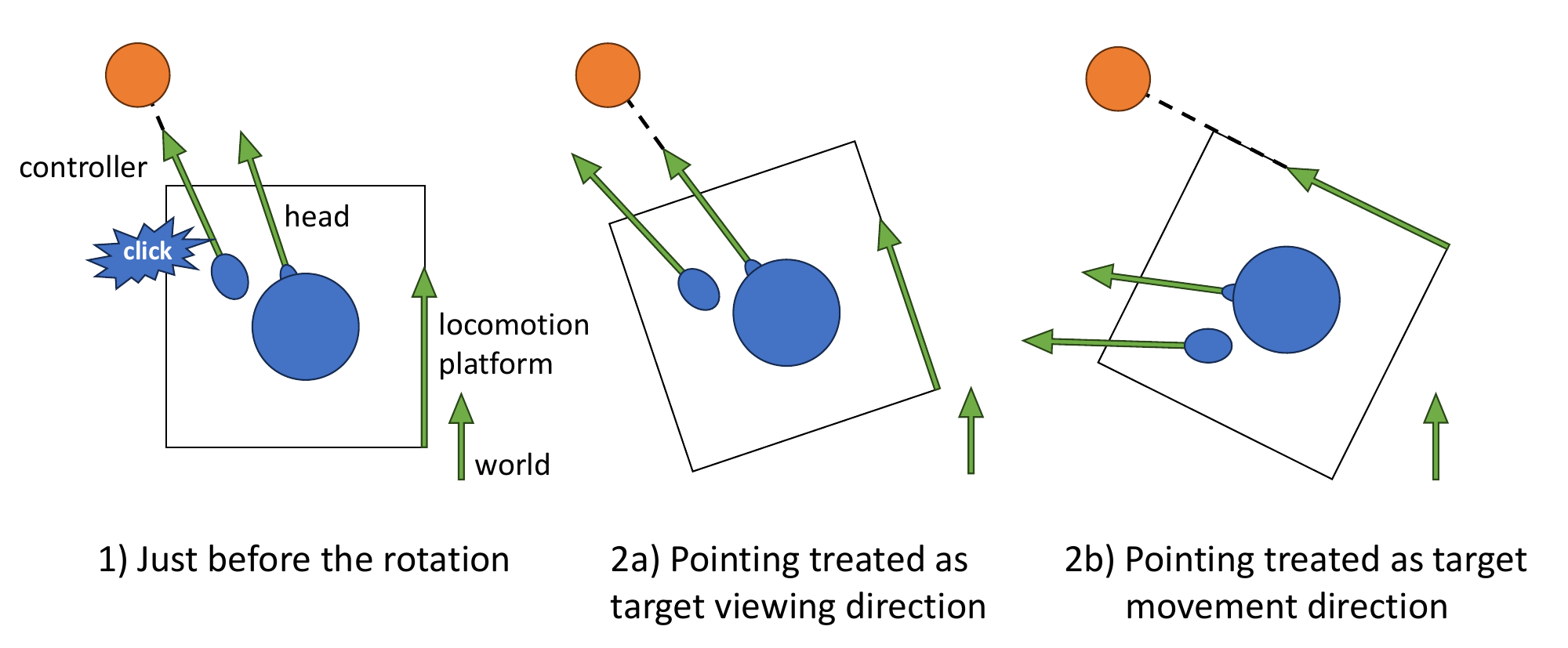}
  \caption{\label{fig_rotationSelection} 
\textbf{Orientation Selection:} \changed{When selecting an orientation by pointing (1), there are at least two ways of interpreting whom's orientation was selected: (2a) the looking/HMD direction's, i.e., the user is looking at the target after the re-orientation or (2b) the locomotion platform's direction, i.e., the user determines the orientation of the virtual pendant of their physical body, which sounds strange but is the way we reorient ourselves in reality when we move our bodies towards a target object and our head's orientation is used to look around. In the common case where the forward direction of the head/HMD and the locomotion platform point in the same direction when rotating, i.e., the user is simply looking forward, both methods behave identically. It is important to add at this point that in our setting, we can assume that the orientation of the motion platform corresponds to that of the user's physical body, as we assume that the user is sitting on a non-rotatable chair. In cases where this is not the case, without the use of additional trackers it remains unclear what the forward direction of the physical body is. In our implementation, we choose (2a).}}
\end{figure}
For the rotation via Orientation Selection, the user indicates a direction with the pose of a tracked handheld controller.
Its forward direction is projected to the ground plane, and when confirmed with a press of the primary index trigger the rotation is performed.
In this approach, the selected direction is interpreted as the user's intended view orientation (cf. \cite{Zielasko2022rotation}), meaning that at the end of the rotation, the user's head will be facing the chosen direction (see Fig.~\ref{fig_rotationSelection}: 2a). 
There are alternative methods, such as applying the selected orientation with respect to the motion\changed{/locomotion} platform\changed{--which describes the virtual coordinate system that the tracked camera is a child of and that is translated and rotated with respect to the global coordinate system of the tracking space to move the user/camera virtually--}instead of the user's head (see Fig.~\ref{fig_rotationSelection}: 2b).
At first glance, this even sounds more useful because it is possible to look at a target and select it to be oriented with respect to the travel direction. 
In other words, the user can choose a target that they can move towards directly. 
In the variant we picked (2a), nothing happens when selecting a target directly in the line of sight. 
However, it quickly becomes apparent that without additional references in the virtual world, such as a virtual body or virtual shoulders indicating the forward orientation, it is challenging to perform such selections (2b) with respect to the virtual body \changed{(locomotion platform)}.
When the user looks in the direction of movement, both methods behave identically because the reference coordinate systems are aligned. 
This is indeed the common case.

Last, it is possible to keep the trigger pressed. 
Then the next rotation will happen with a $0.4~s$ cool-down/delay, following \cite{Farmani2020evaluating, Zielasko2022rotation}.

\subsubsection{Rotation: Snap Turn\label{sec_implementation_snap}}
Since \changed{the rotation is applied only} around the Yaw axis with all methods, the rotation via Snap-Turning only requires a binary input (left/right).
We use two varying inputs here.
With the first option, the user turns left when they point to the left, with respect to their head, and they rotate to the right when pointing to the right \changed{(Pose)}.
Alternatively, with the second option, the direction is determined by the sign of an analog/thumbstick's x-axis on a tracked handheld controller.
In simpler terms, this means the user rotates left when the analog stick is moved to the left and turns right when the stick is moved to the right  \changed{(Thumbstick)}.
The direction is evaluated in the controller's reference space.
When confirmed, in both options the viewpoint is instantly rotated in the respective direction around the global Yaw axis by $11.25^{\circ}$, following \cite{Zielasko2022rotation}.
The confirmation is either triggered by pressing the trigger, or by tilting the analog stick by more than 0, as achieving a \changed{deflection of full magnitude} in the dedicated axis can be challenging with \changed{an analog stick}.
Further, it is possible to lock the confirmation status, i.e., keep pressing. 
Then the next rotation will happen with a $0.4~s$ cool-down/delay.

\subsubsection{Mappings and Conditions\label{sec_implementation_conditions}}
Now we combine the conditions according to the two factors with two levels each.
In the combined methods, the methods that do not rely on Leaning for translation require two tracked controllers, in a naive implementation.
Thus, we decide to specify rotation consistently using the user's dominant hand across the methods since rotation is performed using a handheld controller in all cases, while translation varies in this aspect.
This leads to the following methods:

\begin{table}[h]
\renewcommand{\tabularxcolumn}[1]{m{#1}}
\begin{tabularx}{\columnwidth}{lX}
\protect\icon{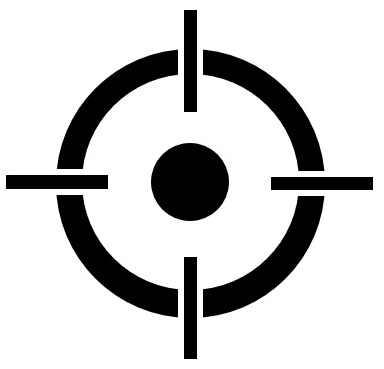}\protect\icon{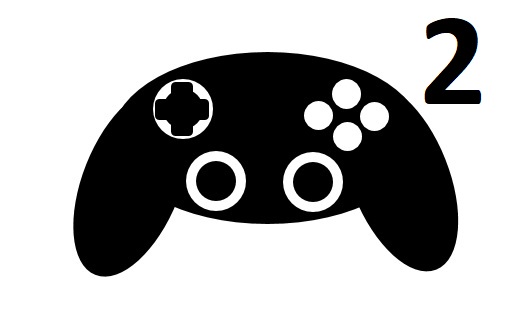} & \textbf{SelecPoint}: Selection-Based Orientation + Pointing-Directed Steering, with 2 handheld controllers \\
\protect\icon{icon_target.jpg}\protect\icon{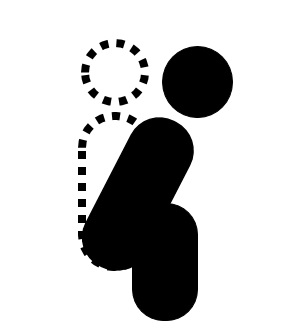} & \textbf{SelecLean}: Selection-Based Orientation + Leaning, with 1 handheld controller \\
\protect\icon{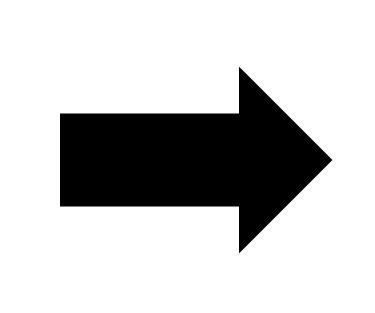}\protect\icon{icon_gamepad_2.jpg} & \textbf{SnapPoint}: Snap Rotation (Pose) + Pointing-Directed Steering, with 2 handheld controllers \\
\protect\icon{icon_pointing.jpg}\protect\icon{icon_leaning2} & \textbf{SnapLean}: Snap Rotation (Pose) + Leaning, with 1 handheld controller
\end{tabularx}
\end{table}

We have previously engaged in discussions regarding the potential challenges arising from the simultaneous utilization of two controllers, each assigned with a distinct task, which may pose a potential impediment to the overall usability of the interfaces.
In light of this concern, we have contemplated whether conflicts in input mappings could be addressed with alternative interfaces. 
Zielasko et al.~\cite{Zielasko2022rotation} conducted a study in which they evaluate an alternative approach for specifying the direction of Snap Turning using \changed{a} thumbstick of \changed{a} controller (see Section~\ref{sec_implementation_snap}). 
Notably, in their evaluation, this alternative method exhibited only a marginal decrease in subjective usability. 
We make the decision to incorporate this alternative method into our deliberations and further investigate its potential applicability within the current framework.

\begin{table}[h]
\renewcommand{\tabularxcolumn}[1]{m{#1}}
\begin{tabularx}{\columnwidth}{lX}
\protect\icon{icon_pointing.jpg}\protect\icon{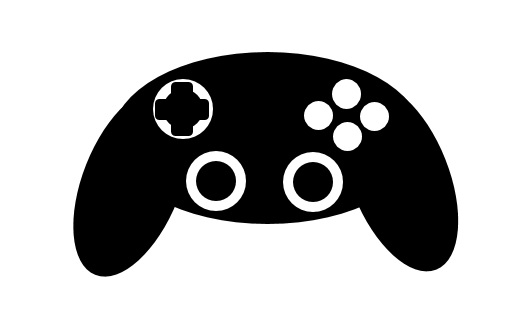} & \textbf{SnapAltPoint}: Snap Rotation (Thumbstick) + Pointing-Directed Steering, with 1 handheld controller
\end{tabularx}
\end{table}

We have considered this solution also as an alternative approach for the first-mentioned condition, \textit{SelecPoint}.
We decided to abandon this variation because choosing an orientation, which is significantly different from specifying a binary direction (left/right) with the thumbstick on a tracked controller while also indicating the direction of travel, felt more challenging \changed{during rapid prototyping} than operating the controllers simultaneously.
With \textit{SnapAltPoint} the interface requires only one controller/hand but translation and rotation are still manipulated by two distinct operations (pointing \& analog stick).
As humans tend to serialize asymmetrical bimanual interactions because of their complexity \cite{Swinnen1991control, Balakrishnan2000Symmetric}, we expect to observe this behavior also here.
Acknowledging this expectation, we implement a last interface that is operated only by pointing with the tracked controller.
Depending on the offset between the forward direction of a tracked handheld controller and the user's forward head direction (HMD forward direction) the user then is either translating into \changed{the controller's} direction or, when this offset exceeds $\pm45^\circ$ Snap Rotating \changed{towards the controller's} direction.
Simultaneous manipulation of translation and rotation is no longer possible for the sake of simple operability.

\begin{table}[h]
\renewcommand{\tabularxcolumn}[1]{m{#1}}
\begin{tabularx}{\columnwidth}{lX}
\protect\icon{icon_pointing.jpg}\protect\icon{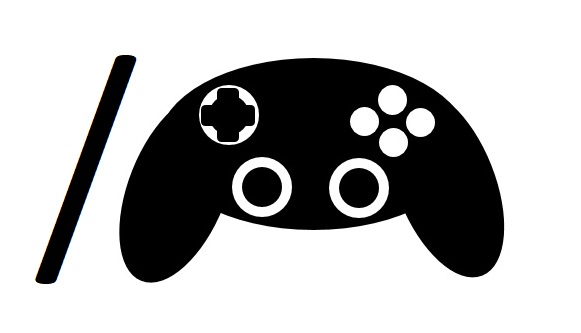} & \textbf{SnapOrPoint}: Snap Rotation \textit{or} Pointing-Directed Steering, with 1 handheld controller
\end{tabularx}
\end{table}

\subsection{Hypotheses\label{sec_hyp}}
While in previous sections we have occasionally expressed different expectations and partly provided rationale, especially regarding bimanual vs. unimanual operation, we systematically formulate and derive hypotheses to answer the research questions in the following section.

We assume that users have greater difficulty controlling translation and rotation simultaneously, especially when this requires both hands to perform different tasks, similar to the challenge of learning to play the drums. 
This should lead to a longer learning curve, more errors, a higher cognitive load, and a lower level of satisfaction when attempting to control both translation and rotation simultaneously.
We anticipate that this will specifically manifest in the three measures: usability (H1), task load (H2), and task performance (H3).
Subsequently, we have divided these hypotheses into directly analyzable sub-hypotheses.
In the sub-hypotheses, we consistently compare pairs of methods or pairs of method groups that are very similar, except in the key factor of handedness.

\subsubsection{Unimanual vs. Bimanual Interaction}
Higher subjective \textbf{usability} of the (unimanual) Leaning interfaces (\textit{SelecLean \& SnapLean}) compared to the bimanual pointing interfaces (\textit{SelecPoint \& SnapPoint}) \textbf{(H1.1)}. 

Higher subjective usability of the unimanual, alternative interface (\textit{SnapAltPoint}) compared to the respective bimanual pointing interface (\textit{SnapPoint}) \textbf{(H1.2)}. 

Higher subjective usability of the unimanual, alternative interface (\textit{SnapOrPoint}) compared to the respective bimanual pointing interface (\textit{SnapPoint}) \textbf{(H1.3)}.


We expect that the above-formulated sub-hypotheses are mirrored by our measures of \textbf{task load} \textbf{ (H2.1-H2.3)}.

Furthermore, we anticipate observing a similar pattern for the last measure, \textbf{task performance} \textbf{(H3.1-H3.2)}, but with an exception in (H3.3).
Based on our experiences from previous studies in this field, we generally assumed that participants would perceive bimanual interfaces as more complex. 
However, we expect that participants also quickly adapt to the circumstances. 
Additionally, with \textit{SnapOrPoint} (the unimanual interface) rotation and translation must be performed sequentially.
\textit{SnapPoint} allows but does not enforce sequential execution.
This means it is already faster to occasionally perform rotation and translation in parallel with the latter.
Thus, we anticipate that the performance measure is the opposite, i.e., worse with \textit{SnapOrPoint} (unimanual) than with \textit{SnapPoint} (bimanual) \textbf{(H3.3)}.

Last, we want to add an additional hypothesis for task performance, which revisits the previously mentioned circumstance that led to H3.3.
For the reasons stated, we expect worse task performance, but the variance in performance with \textit{SnapOrPoint} should be smaller than with \textit{SnapPoint} because the ostensibly simpler interface should be more invariant across user groups/experiences \textbf{(H3.4)}.

\subsubsection{Orientation Selection vs. Snap Rotation}
Since Snap Rotation allows only for discrete jumps of 11.25$^\circ$, we expect that methods using Snap Rotation lead to less \textbf{accurate} movement or maneuvering than methods utilizing Orientation Selection \textbf{(H4)}.
On the other hand, Orientation Selection allows for much quicker maneuvers, such as rotating 180 degrees, which should lead to better \textbf{performance} \textbf{(H5)}.
However, this advantage gained through skipping should also result in poorer path integration or \textbf{orientation}. 
Although this effect is only present at a lower cognitive level in the prior work of Zielasko et al. \cite{Zielasko2022rotation}, in isolation from translation, and ultimately has no verifiable impact on overall orientation performance, we initially assume that we can observe this effect here \textbf{(H6)}.

\subsection{Study Design}
To evaluate our hypotheses we conducted a user study with a within-subjects design.
The study received approval from the local ethics board.
In this study, participants (see Section~\ref{participants}) were presented with a primed search task in a maze. 
In addition, there are elements that require accurate maneuvering and the subjects were asked to remember the position of some landmarks on their route (dual-task design). 
The details are listed in the following. 

\subsubsection{Procedure}
At the outset, participants were instructed to review a study information document, which detailed their roles in the study, the potential experience of adverse symptoms, and the handling of their data. 
Subsequently, they provided their informed consent by signing a declaration.
Each participant was assigned a unique random ID, known only to them and the experimenter. 
This ID was used to pseudonymize their data.
Following this initial process, participants were asked to complete a demographic and prior experiences questionnaire. 
They were then introduced to the hardware (see Section~\ref{apparatus}). 
The interpupillary distance of the participants was measured and adjusted accordingly.
Participants were given an explanation of the tasks (see Section~\ref{task}) and a general overview of the procedure for all six locomotion interfaces (see Section~\ref{interfaces}). 
To eliminate learning biases, the six travel interfaces were tested in varying sequences using a balanced Latin Square design paired with different maze configurations (see Section~\ref{environments}).
Afterward, the first travel interface was verbally described. 

Participants were given the opportunity to practice in a training room different from the actual study environment. 
\changed{The training environment was a large rectangular room ($15~m \times 15~m$) enclosed by walls on all sides. In the center, there was a large green box, and distributed throughout the room were 4 or 5 objects, including the chess piece and billiard ball that were also sought in the labyrinth.
The task was simply to try out the method until one felt comfortable with it.
They did this for each condition and immediately before each of the conditions.}

An experimenter was present throughout the procedure to address any questions or concerns.
Once participants felt confident in using the interface, the actual study task commenced. 
They were virtually positioned at the starting point of the course and proceeded to follow it to its conclusion.
After each run with a specific locomotion method, participants took a break, sat in front of a laptop to draw objects related to the orientation task, and completed a questionnaire specific to the interface.
Subsequently, the experimenter inquired about the participants' well-being. 
If, following a joint evaluation, it was decided to continue with the study, the experimenter explained the next locomotion interface. 
This process was repeated for all six interfaces.
Finally, participants had the opportunity to ask questions about the study and its purpose, which were answered by the experimenter.

\subsubsection{Tasks \& Measures\label{task}}
\begin{figure}
  \centering
  \includegraphics[width=\columnwidth]{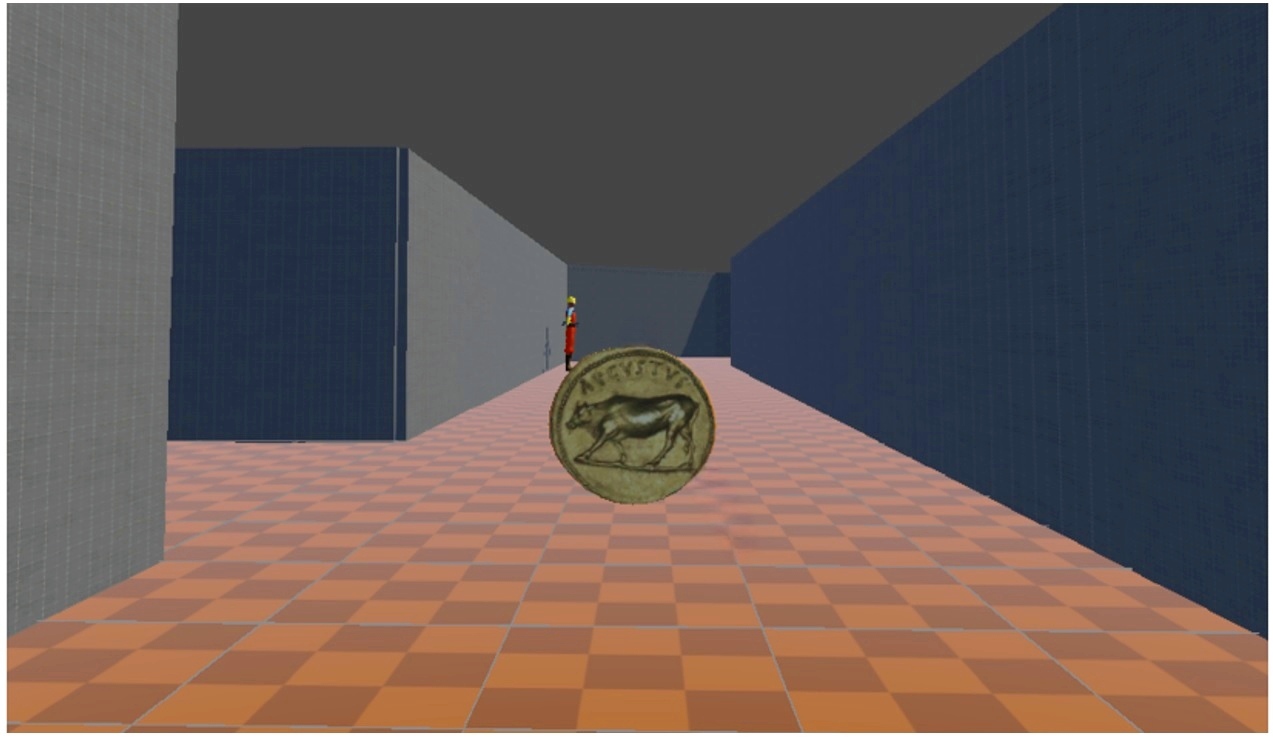}
  \caption{\label{maze} Ego-perspective into the virtual environment. In the foreground the current target coin is visible. In the background, you can discern one of the items to be remembered, a construction worker.}
\end{figure}

The study followed a dual-task design, which is common to reveal effects on cognitive load in locomotion research.
Further, it emulates that travel in most applications is not the main purpose but a tool.
In this study, we aimed to explore an extended utilization of this construct. 
Measuring spatial orientation, especially in the context of locomotion, has proven to be quite challenging in the past, leading to the existence of numerous approaches. 
Frequently repeated rapid pointing tasks have recently shown to be relatively successful in revealing effects~\cite{Zielasko2022rotation, Adhikari2022hyperjump} but pose a significant source of interference when integrated into a travel task, as the actual locomotion must be repeatedly and substantially interrupted~\cite{Adhikari2022hyperjump}. 
On the other hand, asking participants only after the search tasks provides too few data points to overcome statistical noise.
Another approach commonly used to measure spatial orientation is having participants draw maps, paths, or similar elements. 
One major challenge here is quantifying the results, as different individuals store spatial data in their brains differently. 
Remembering and visualizing such data may yield fundamentally different results, making it difficult to draw conclusions about the quality of navigation. 
For these reasons, we are attempting an experimental approach in this work.
We provide participants with objects to memorize as they navigate through the mazes. 
After completing a run, we ask them to draw the locations of these objects on a grid-based map that shows the outline of the maze.
It is not part of the task to identify the objects individually.
Every grid cell represents $10~m$ x $10~m$ and the errors are evaluated using a Manhattan distance.
However, we anticipate that the analysis of these maps may not yield sufficient data for meaningful statistical comparisons given our chosen sample size. 
Instead, we assume that if a locomotion method has a more pronounced negative impact on \textbf{orientation} than another, this should be reflected in the subjective but standardized task load measure (NASA-TLX \cite{hart1988development}). 
This is because the effort to memorize interferes with the mental load required to operate the interfaces.
The major limitation of this method is evidently that an increased task load can only serve as an indication of poorer or negatively influenced orientation, as it may intertwine with other factors (see Section~\ref{sec_discussion}).

In general, participants were informed that their primary objective was to complete the naive search task as quickly as possible, and thus task completion time was our main \textbf{performance} measure.
A clear path for following the route in an otherwise branching maze is established by sequentially introducing coins that must be collected without exception (see Figure~\ref{maze}).
The successful collection of a coin is acknowledged by corresponding auditory feedback, followed by the immediate appearance of the next coin.
As additional goals for quickly completing a level, participants were encouraged to pass through the coins as centrally as possible and to memorize the positions of the aforementioned three objects as they passed by.
Thus, the distance to the center of the coins served as our measure of \textbf{accuracy}.
Coins are always collected whenever they are passed, regardless of the distance, to eliminate additional sources of interference. 
Participants were aware of this.
However, none of the participants markedly ignored the objective to be accurate.

In addition to the described measurements, we also assessed \textbf{usability} using the System Usability Scale (SUS) \cite{brooke1996sus}, subjective orientation using a single-item 4-point Likert scale, and well-being using a 10-point variation (cf. \cite{zielasko2018dynamic}) of the Fast Motion Sickness Scale (FMS) \cite{Keshavarz2011fms} as a \textbf{control variable}.

\subsubsection{Virtual Environments\label{environments}}
\begin{figure}
  \centering
  \includegraphics[width=\columnwidth]{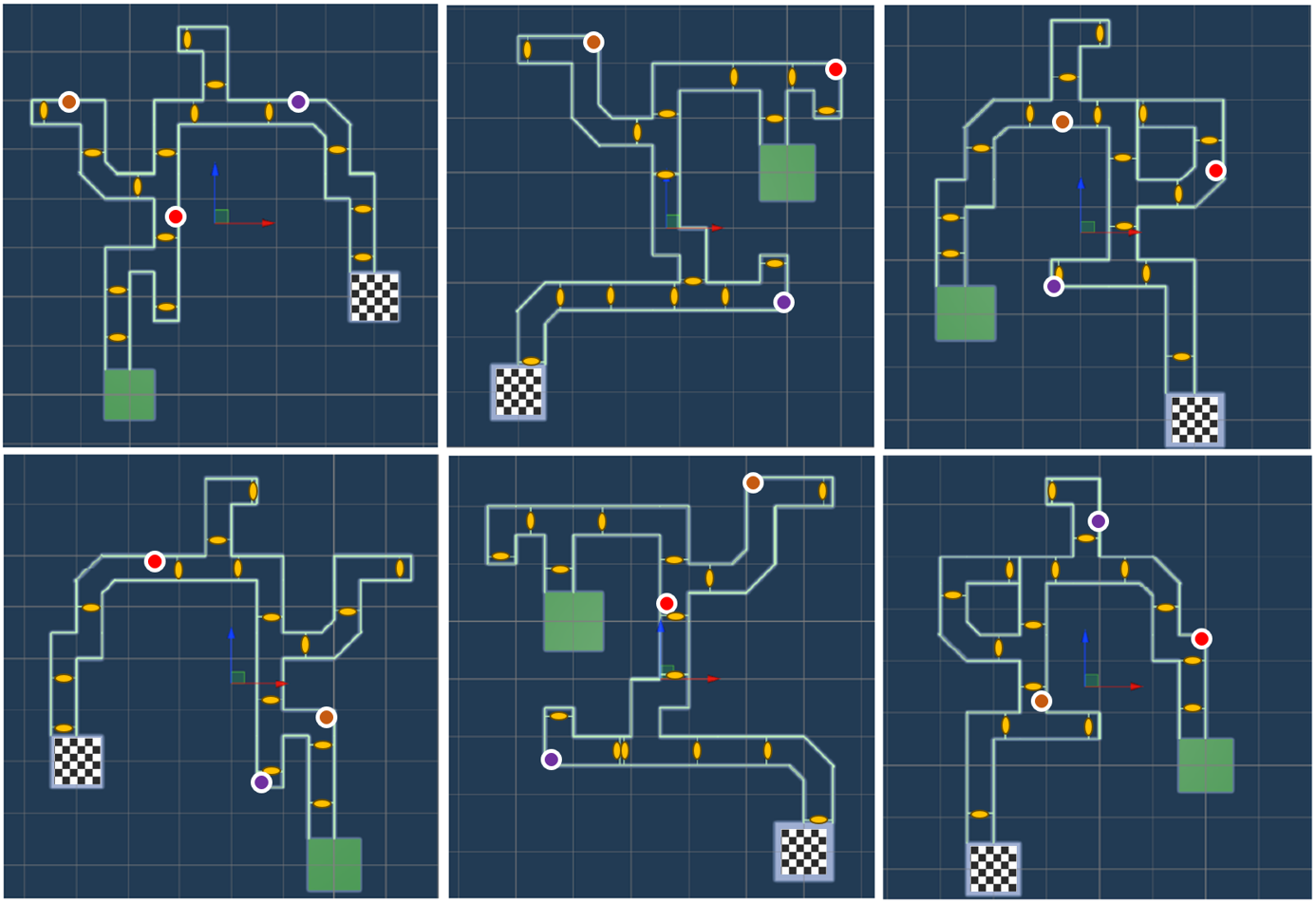}
  \caption{\label{mazes_measurements}Measurement of the six labyrinth courses. Superimposed are the positions of the memorizable objects (circles) and the positions of the coins (ellipses). The size of the overlays does not accurately represent their actual size.}
\end{figure}

Three courses were designed for the experiment, each of which was mirrored, resulting in a total of six courses.
Despite varying the courses, we aimed to make them comparable based on their core characteristics. 
Each course consists of three dead-end paths and eight curves, with four right turns and four left turns to navigate. 
Due to the dead-end paths, some curves need to be traversed twice. 
The course boundaries are defined by walls and the width of the corridor is 4.5 meters. 
The length of each course ranges from 230 to 245 meters. 
Participants complete an entire course during each trial, which means they traverse each dead-end path. 
Due to the dead-end paths, virtual rotations of 180$^\circ$ are required.
The angles of the curves are either 45$^\circ$ or 90$^\circ$. 
A curve that involves a total rotation of 90$^\circ$ can be divided into two 45$^\circ$ angles. 
Each course includes two curves of this type, and one of them is traversed twice as it leads to a dead-end path. 
As a result, participants need to rotate 180$^\circ$ three times, 45$^\circ$ five times, and 90$^\circ$ thirteen times within one designed course.
Consequently, in a designed course, there is one more 45$^\circ$ rotation either to the left or right, whereas in the mirrored course, there is one more 45$^\circ$ rotation in the opposite direction.
Finally, every maze contains the same set of three objects: a large chess knight, a large cue ball, and a construction worker.
Their placement and the dimensions of the six courses are depicted in Figure~\ref{mazes_measurements}.

\subsubsection{Apparatus\label{apparatus}}
The study application was conducted on an Oculus Quest (2019). 
\changed{In addition two tracked Oculus Quest hand-held controllers were used}.
Participants were seated in a chair without swivel capability. 
The chair was positioned in the center of the room to maintain a safe distance from other objects in the room. 
Additionally, a laptop was used for participants to complete questionnaires.
The study application was developed in Unity 2021.3.4f1, utilizing the Oculus XR Plugin 3.0.2.

\subsubsection{Participants\label{participants}}
In the study, a total of \changed{42 participants} took part, consisting of \changed{21 women} and 21 men.
Participants were reached through distribution channels at the university and the experimenter's professional and personal networks.
For their participation, they were compensated with 10€.
The average age of the participants was 24.9 years, with a standard deviation of \changed{5.5 years} (\changed{median 24}, ranging from \changed{19 to 55}). 
In the age group 18-28, there were \changed{18 women} and 18 men. 
In the 28+ age group, there \changed{were three} woman and three men.
\changed{29} participants reported having prior experience with VR, while \changed{13} participants reported having no prior experience. 
Among the \changed{29} participants with prior VR experience, there were 14 men and \changed{15 women}. 
Among the \changed{13} participants with no prior VR experience, there were 7 men and \changed{6} women. 
Sixteen participants indicated that they regularly played 3D video games, whereas \changed{26 participants} stated that they did not play 3D video games regularly. 
Among those who regularly played 3D video games, there were 14 men and 2 women, while among those who did not play 3D video games regularly, there were 7 men and \changed{19 women}.

\section{Results\label{sec_results}}
\begin{table*}
\caption{Pairwise (Condition A vs. B) inferential statistical analysis of the hypotheses, where Hyp. specifies the hypothesis number, Dir. denotes the direction of the hypothesis, DV the dependent variable, followed by the t or Z statistics. 
\changed{The Z statistic is given in case the requirements of the paired t-test were not met and a non-parametric test was performed.
These cases are identifiable with the SDs, dfs not specified and a significant Shapiro-Wilk test indicating non-normally distributed residuals.
In cases where the opposite direction of the hypothesized effect was significant, we can reject our hypothesis and denote this with an \textbf{r}ejected in the last column.
The last column further codes \textbf{c}onfirmed, not confirmed (\textbf{-}).}
Note that Lean = Avg(SelecLean, \changed{SnapLean}), Point = Avg(SelecPoint, SnapPoint), Snap = Avg(SnapPoint, SnapLean), and Selec = Avg(SelecPoint, SelecLean) are derived from the per participant average of the respective conditions.\label{tab_inference}}
\resizebox{\textwidth}{!}{%
\begin{tabular}{cclrrclrrcrrrrrl}
\textbf{Pair} & \textbf{Hyp.} &
  \textbf{Condition A} &
  M~/~Mdn &
  SD &
  \textbf{Dir.} &
  \textbf{Condition B} &
  M~/~Mdn &
  SD &
  \textbf{DV} & \textbf{S.-Wilk} &
  \textbf{t~/~Z} &
  \textbf{df} &
  \textbf{p} &
  \textbf{d~/~r} & \\
\rowcolor[HTML]{EFEFEF} 1  & 1.1 & Lean         & 74.6  & 16.4  & \textgreater{} & Point     & 80.0     & 12.5      & SUS      & .45                        & -2.05     & 41    & .98  & .32  & r \\
                        2  & 1.2 & SnapAltPoint & 83.2  & 15.6  & \textgreater{} & SnapPoint & 74.6     & 19.7      & SUS      & .08                        & 3.05      & 41    & \textbf{.002}             & .47  & c \\
\rowcolor[HTML]{EFEFEF} 3  & 1.3 & SnapOrPoint  & 67.4  & 22.3  & \textgreater{} & SnapPoint & 74.6     & 19.7      & SUS      & .09                        & -1.8      & 41    & .96                       & .28  & r \\
                        4  & 2.1 & Lean         & 24.8  & -     & \textless{}    & Point     & 14.8     & -         & TLX      & \textbf{\textless.001}     & -3.48     & -     & \textgreater.99           & .54  & r \\
\rowcolor[HTML]{EFEFEF} 5  & 2.2 & SnapAltPoint & 12.9  & -     & \textless{}    & SnapPoint & 15.8     & -         & TLX      & \textbf{\textless.001}     & 1.14      & -     & .13                       & .18  & - \\
                        6  & 2.3 & SnapOrPoint  & 21.6  & 17.9  & \textless{}    & SnapPoint & 21.5     & 19.7      & TLX      & .03                        & .05       & 41    & .52                       & .01  & - \\
\rowcolor[HTML]{EFEFEF} 7  & 3.1 & Lean         & 141   & 21.6  & \textless{}    & Point     & 127      & 23.2      & Time     & .29                        & 3.62      & 41    & \textgreater.99           & .56  & \changed{r} \\
                        8  & 3.2 & SnapAltPoint & 126   & 15.8  & \textless{}    & SnapPoint & 135      & 28.3      & Time     & .25                        & -2.41     & 41    & \textbf{.01}              & .37  & c \\
\rowcolor[HTML]{EFEFEF} 9  & 3.3 & SnapOrPoint  & 161   & 25.3  & \textgreater{} & SnapPoint & 135      & 28.3      & Time     & .95                        & 4.57      & 41    & \textbf{\textless{}.001}  & .71  & c \\
                        10 & 4   & Snap         & .084  & .03   & \textless{}    & Selec     & .081     & .026      & Accuracy & .65                        & .72       & 41    & .76                       & .11  & - \\
\rowcolor[HTML]{EFEFEF} 11 & 5   & Snap         & 143   & 22.3  & \textgreater{} & Selec     & 125      & 18.7      & Time     & .56                        & 6.38      & 41    & \textbf{\textless{}.001}  & .99  & c \\
                        12 & 6.1 & Snap         & 22.3  & -     & \textless{}    & Selec     & 18.8     & -         & TLX      & \textbf{\textless.001}     & -.88      & -     & .81                       & .14  & - \\
\rowcolor[HTML]{EFEFEF} 13 & 6.2 & Snap         & 1.63  & 1.3   & \textless{}    & Selec     & 1.52     & 1.1       & Map      & .062                       & .50       & 41    & .69                       & .08 & -
\end{tabular}}
\end{table*}
\begin{figure*}
    \centering
    \begin{minipage}{.49\textwidth}
        \includegraphics[width = \textwidth]{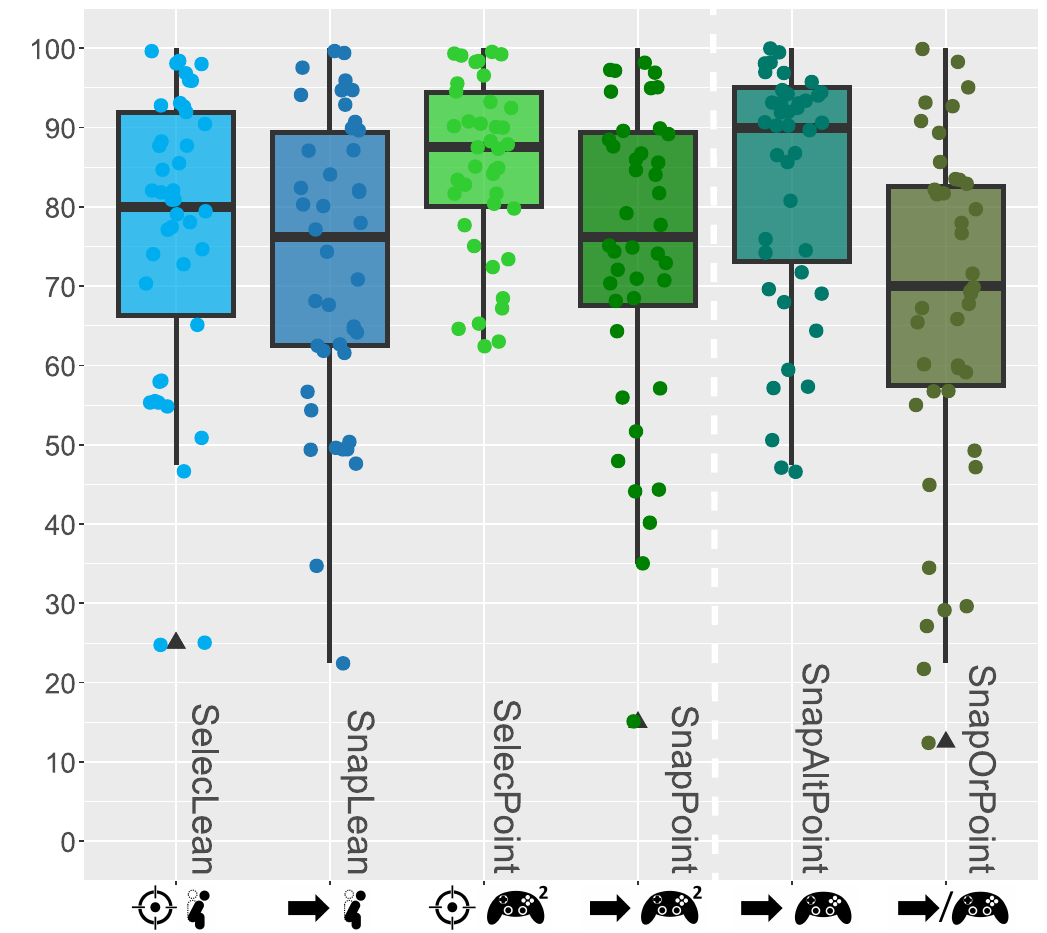}
        \caption{SUS\label{img_result_sus}}  
    \end{minipage}
    \begin{minipage}{.49\textwidth}
        \includegraphics[width = \textwidth]{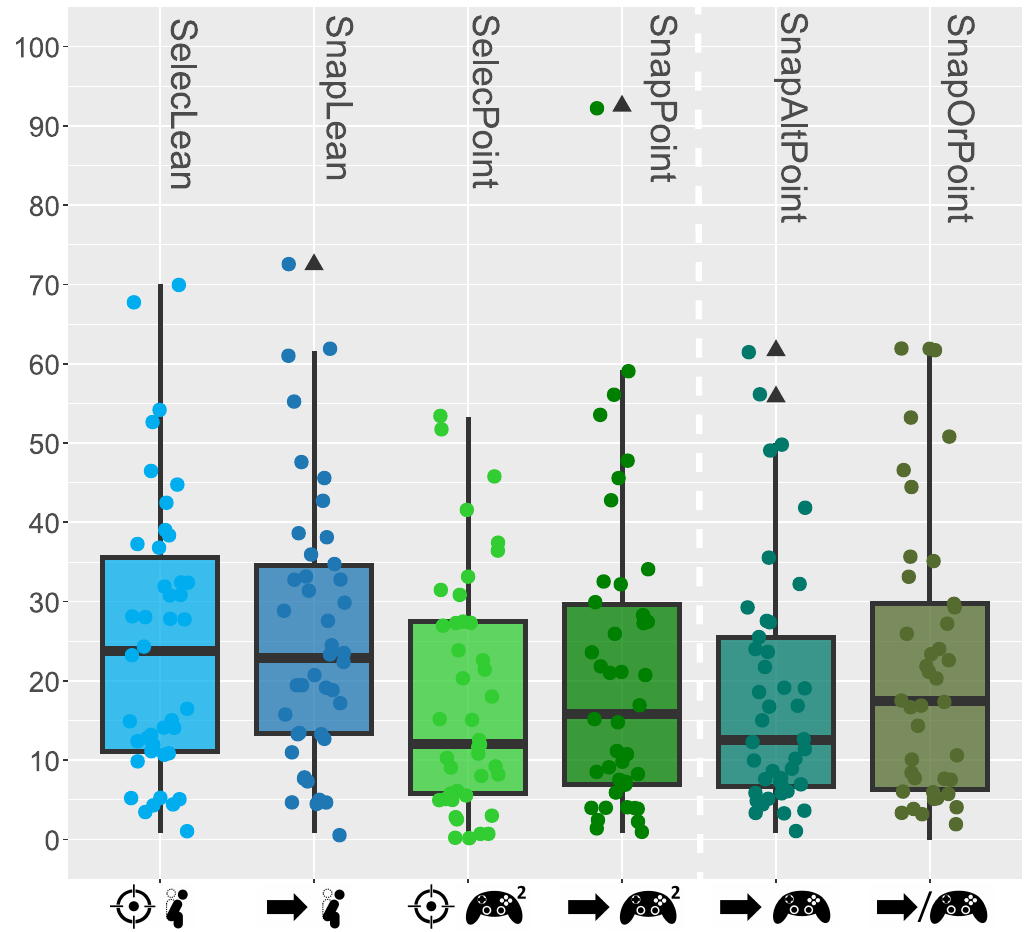}
        \caption{NASA TLX\label{img_result_tlx}} 
    \end{minipage}
\end{figure*}
\begin{figure*}
    \centering
       \begin{minipage}{.33\textwidth}
        \includegraphics[width = \textwidth]{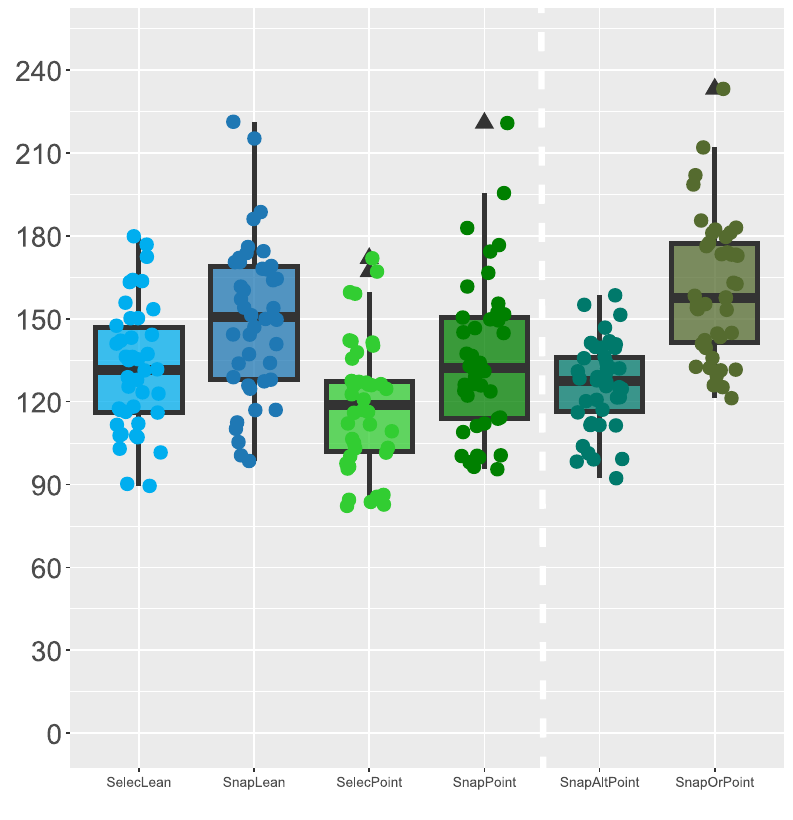}
        \caption{Completion Time in Seconds\label{img_result_time}}   
    \end{minipage}
    \begin{minipage}{.33\textwidth}
        \includegraphics[width = \textwidth]{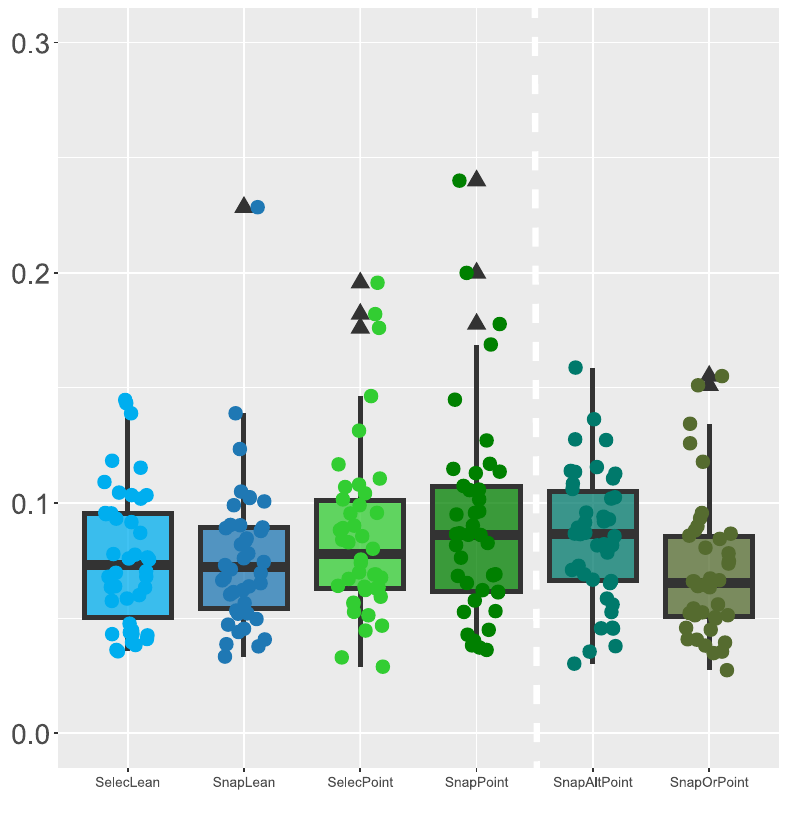}
        \caption{Accuracy in Meter\label{img_result_accuracy}} 
    \end{minipage}
    \begin{minipage}{.33\textwidth}
        \includegraphics[width = \textwidth]{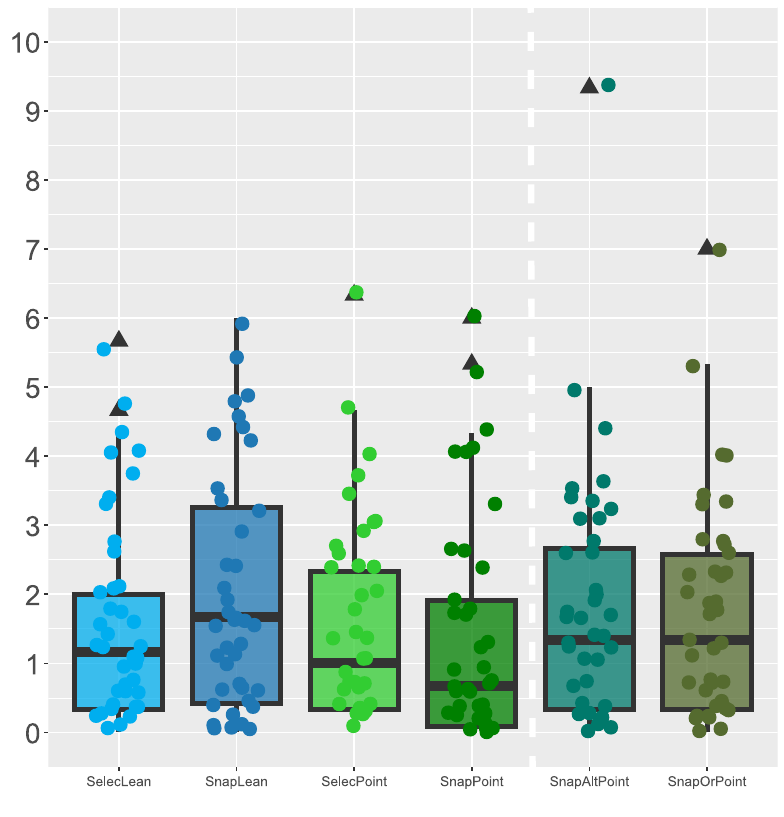}
        \caption{Map Drawing Error in Manhattan Distance\label{img_result_mapErrors}}  
    \end{minipage}
\end{figure*}
\begin{table*}
\caption{Descriptive statistics (M = mean, SD = standard deviation) of all measures. The MapDrawing Error measures the average placement error per object in Manhattan distance.\label{tab_descr}}
\centering
\scriptsize
\begin{tabular}{lrrrrrrrrrrrr}
 &
  \multicolumn{2}{r}{\textbf{SelecLean}} &
  \multicolumn{2}{r}{\textbf{SnapLean}} &
  \multicolumn{2}{r}{\textbf{SelecPoint}} &
  \multicolumn{2}{r}{\textbf{SnapPoint}} &
  \multicolumn{2}{r}{\textbf{SnapAltPoint}} &
  \multicolumn{2}{r}{\textbf{SnapOrPoint}} \\
                  & M     & SD   & M     & SD   & M     & SD   & M     & SD   & M     & SD   & M     & SD   \\
\rowcolor[HTML]{EFEFEF}\textbf{SUS}               & 76.5  & 18.8 & 72.7  & 19.0 & 85.4  & 11.3 & 74.6  & 19.7 & 83.2  & 15.6 & 67.4  & 22.3 \\
\textbf{NASA TLX}          & 24.8  & 17.5 & 25.7  & 17.2 & 17.7  & 14.8 & 21.5  & 19.7 & 18.2  & 15.4 & 21.6  & 17.9 \\
\rowcolor[HTML]{EFEFEF}~~$\circ$ Mental Demand   & 29.4  & 23.2 & 29.9  & 25.7 & 21.8  & 20.9 & 25.7  & 24.6 & 21.1 & 19.8 & 27.4  & 26.0 \\
~~$\circ$ Physcial Demand & 27.1  & 21.6 & 29.1  & 22.2 & 9.5   & 14.1 & 14.4  & 21.9 & 8.0  & 15.7 & 11.9  & 18.9 \\
\rowcolor[HTML]{EFEFEF}~~$\circ$ Temporal Demand & 19.1  & 23.1 & 18.8  & 23.0 & 15.0  & 25.7 & 14.2  & 19.7 & 16.2  & 23.3 & 15.5  & 20.9 \\
~~$\circ$ Performance     & 31.2  & 26.7 & 30.5  & 23.9 & 26.8  & 22.7 & 29.2  & 26.4 & 27.1  & 22.7 & 31.8  & 24.3 \\
\rowcolor[HTML]{EFEFEF}~~$\circ$ Effort          & 26.1  & 23.4 & 29.8  & 22.1 & 20.4  & 20.0 & 26.6  & 26.2 & 21.8  & 21.9 & 24.3  & 23.0 \\
~~$\circ$ Frustration     & 16.1  & 23.7 & 16.3  & 19.1 & 12.6   & 18.0 & 18.9  & 24.0 & 14.9  & 21.5 & 18.8  & 24.1 \\
\rowcolor[HTML]{EFEFEF}\textbf{Time} {[}s{]}      & 132 & 22.8 & 150 & 28.3 & 118 & 23.2 & 135 & 28.3 & 126 & 15.8 & 161 & 25.3 \\
\textbf{Accuracy} {[}m{]}  & .077  & .03 & .077  & .034 & .086  & .037 & .091  & .044 & .085  & .029 & .071  & .031 \\
\rowcolor[HTML]{EFEFEF}\textbf{MapDrawing Error}        & 1.57  & 1.47 & 1.91  & 1.74 & 1.47  & 1.49 & 1.35  & 1.61 & 1.77  & 1.81 & 1.68  & 1.57 \\  
\textbf{FMS} {[}0, 10{]}   & 1.55  & 2.03 & 1.70  & 2.32 & 1.17   & 1.76 & 1.43   & 2.01 & 1.43   & 2.01 & 1.38  & 1.93 \\
\end{tabular}
\end{table*}

The descriptive statistics of our measures are listed in Table~\ref{tab_descr} and visualized in Figures~\ref{img_result_sus}-\ref{img_result_mapErrors}.
In the following, we will systematically evaluate our hypothesis (see Section~\ref{sec_hyp}) and then have an exploratory look at our data in Section~\ref{sec_results_explore}. 

To validate our hypotheses we use one-sided Paired Sample t-Tests.
Whenever the data violated the assumption of being normally distributed, determined by a Shapiro-Wilk test \cite{ShapiroWilk65}, or in the presence of outliers, we use a non-parametric Wilcoxon signed-rank test instead.
We report the significance level at $0.05$ \cite{fisher1956statistical}.
Effect sizes are assessed according to Cohen \cite{cohen2013statistical}.
The complete inferential statistics are listed by hypotheses in Table~\ref{tab_inference}.

In our study, we proposed three main hypotheses (H1, H2, H3) with three sub-hypotheses each, focusing on the negative impact of bimanual interaction compared to unimanual interaction. 
These hypotheses were related to usability (H1), task load (H2), and performance (completion time, H3).
Each sub-hypotheses predicts the pairwise relation of two subgroups of methods that vary in their handiness but are otherwise as similar as possible regarding other factors (see Section~\ref{sec_hyp}).
Since the common difference between all sub-hypotheses then appears to be in summary related to handedness, a measurable difference in each pair would serve as clear evidence for the negative influence of bimanual interaction independent of other factors.
However, we do not observe such a consistent pattern in any of the three main hypotheses. 
In total, only three of the nine sub-hypotheses displayed the expected effect (see Table~\ref{tab_inference}, Hyp. 1.2, 3.2., and 3.3). 
Surprisingly, \changed{four} sub-hypotheses even demonstrated a significant effect in the opposite direction (see Table~\ref{tab_inference}, Hyp. 1.1, 1.3, 2.1 \changed{and 3.1}).
Furthermore, in \textbf{H3.3} as the exceptional case, we suspected that the unimanual method is actually slower than bimanual (\textit{SnapOrPoint} vs. \textit{SnapPoint}) due to another factor, and this suspicion is under the three confirmed ones.

The assumed factor behind this last hypothesis H3.3 was that experienced users in the unimanual \textit{SnapOrPoint} are excessively constrained and thus as slow as inexperienced users. 
As a direct consequence, we expected in \textbf{H3.4} that the variance between participants in performance with \textit{SnapOrPoint} \changed{($\sigma^2 = 641.2$)} is smaller than with \textit{SnapPoint} \changed{($\sigma^2 = 802.7$)}.
A Hartley's Fmax test reveals that this is not the case and we can assume homogeneity of variance $F_{max} = \changed{1.58} < 2.07$, which does mean that we cannot confirm \textbf{H3.4}.

Overall, we cannot confirm \textbf{H1-H3}. 
Furthermore, we tend to hypothesize that asynchronous bimanual interaction indeed does not have a negative impact on our results and the few differences we see might be artifacts from interactions with other modalities (see Section~\ref{sec_results_explore}).

The remaining hypotheses we formulated, \textbf{H4-6}, revolve around the comparison of Orientation Selection to the more common Snap Rotation method.
First, we predicted a lower accuracy with selection-based interfaces \textbf{H4}, which we cannot confirm (\changed{$p = .76$}, see the details in Table~\ref{tab_inference}).
Second, we predicted selection-based interfaces to be faster in completion time \changed{\textbf{H5}}, which we can confirm ($p < .001$) with a strong effect size (\changed{$d = .99$}).
Third, we expected a more challenging orientation with selection-based interfaces, manifesting in an increased task load (\textbf{H6.1}) and a higher error in map drawing (\textbf{H6.2}).
Both cannot be confirmed (\changed{$p = .81$ and $p = .69$}).

\subsection{Exploratory Analysis\label{sec_results_explore}}
As discussed earlier, we lack preliminary results to make predictions regarding the comparison between Leaning and Pointing-Directed Steering. 
\changed{Therefore, in this exploratory analysis, we examine our $2 \times 2$ design with factors \textit{Translation} (Leaning, Pointing-directed Steering) and \textit{Rotation} (Snap Rotation, Rotation Selection).
This means we leave the experimental and hybrid methods \textit{SnapAltPoint} and \textit{SnapOrPoint} out for this part of the analysis.}

When focusing on handedness, no \changed{significant differences were observed}. 
However, when comparing Leaning to Pointing, differences do become apparent.

\changed{A two-way repeated measures ANOVA was conducted that examined the effect of \textit{Translation} and \textit{Rotation} on the \textbf{SUS} score. 
There was no statistically significant interaction between the effects found, $F(1, 42) = 3.52$, $p = .068$.
The results indicated a significant main effect for \textit{Translation}, $F(1, 42) = 4.19$, $p = .047$; and a significant main effect for \textit{Rotation}, $F(1, 42) = 9.30$, $p = .004$.}

\changed{\textbf{NASA TLX} scores violate the ANOVA's assumption to be normally distributed and without significant outlier.
Therefore, we used non-parametric Wilcoxon signed-rank tests to investigate simple effects of \textit{Translation}.
\textit{SelecLean} led to a significant higher NASA TLX score then \textit{SelecPoint} $Z = -2.81$, $p = .005$, $r = .43$.
\textit{SnapLean} led to a significant higher NASA TLX score then \textit{SnapPoint}
$Z = -2.39$, $p = .017$, $r = .37$.}

\changed{A two-way repeated measures ANOVA was conducted that examined the effect of \textit{Translation} and \textit{Rotation} on the \textbf{Time}. 
There was no statistically significant interaction between the effects found, $F(1, 42) = .28$, $p = .87$.
The results indicated a significant main effect for \textit{Translation}, $F(1, 42) = 13.1$, $p < .001$; and a significant main effect for \textit{Rotation}, $F(1, 42) = 40.7$, $p < .001$.}

\changed{A two-way repeated measures ANOVA was conducted that examined the effect of \textit{Translation} and \textit{Rotation} on the \textbf{Accuracy}. 
There was no statistically significant interaction between the effects found, $F(1, 42) = .33$, $p = .57$.
The results indicated a significant main effect for \textit{Translation}, $F(1, 42) = 4.15$, $p = .048$; and no significant main effect for \textit{Rotation}, $F(1, 42) = .33$, $p = .57$.}

\changed{A two-way repeated measures ANOVA was conducted that examined the effect of \textit{Translation} and \textit{Rotation} on the \textbf{MapDrawing Error}. 
There was no statistically significant interaction between the effects found, $F(1, 42) = 1.15$, $p = .29$.
The results indicated no significant main effect for \textit{Translation}, $F(1, 42) = 2.07$, $p = .16$; and no significant main effect for \textit{Rotation}, $F(1, 42) = .25$, $p = .62$.}

\changed{All revealed effects of \textit{Translation} are at the expense of the Human-Joystick method.}


\changed{When}, we assign an adjective rating to each interface based on their mean SUS scores, following the classification by Bangor et al. \cite{Bangor2009}, SnapOrPoint \changed{is rated \textit{OK}, SelecLean, SnapPoint, and SnapLean \textit{good}, and SnapAltPoint and SelecPoint \textit{excellent}}.

\changed{Finally, related work repeatedly finds mixed results regarding gender differences in spatial orientation \cite{Coluccia2004, Munion2019gender}. 
Therefore, we compare the map errors and NASA-TLX scores for both groups.
We do not find a statistically significant difference between map errors with an independent t-test ($t(40) = -.76$, $p = .45$, $d=.24$), nor do we for the NASA-TLX scores with a Mann–Whitney U test ($U = 209$, $Z = -.28$, $p = .78$, $r = .043$).}

\section{Discussion\label{sec_discussion}}
Before we delve into specific aspects of our research questions in the following sections, we would like to discuss some general observations.
First, it should be noted that the sickness scores remained low throughout the experiment (averaging below 1.5 on a scale of 0-10), despite participants continuously navigating through the mazes. 
This initially confirms the motivation to blend discretely with continuous movement components to create less discomforting interfaces. 
However, we do not have a baseline condition to compare to, and due to potential carryover effects, we refrain from delving further into the interpretation of the sickness results.

Furthermore, it is noteworthy that \textit{SnapAltPoint} (Pointing-Directed Steering with Snap Turning via the thumbstick of the same controller) stands out as the best Snap Turning interface across all metrics. 
\changed{This is probably not related to one-handed operability. 
For instance, \textit{SelecPoint}, which is likely the best interface across all metrics,
uses both hands.}
An interesting difference to the bimanual \textit{SnapAltPoint} alternatives is that in \textit{SelecPoint} both hands fulfill the same basic task: pointing in the same way, albeit for different purposes (translation vs. rotation).

\subsection{Uni vs. Bimanual Interaction\label{sec_disc_bimanual}}
The \changed{corpus} of related work investigating the impact of bimanual vs. unimanual interaction in 3D user interaction is small.
The underlying mechanisms in motor control are complex, much of it comes down to the nature of the task and how it is mentally integrated \cite{Talvas2014bimanualSurvey, Owen2005gets, obhi2004bimanual}.
This makes it particularly challenging to extrapolate results from reality to unfamiliar 3D user interfaces.
Our intuition and experience of novice (non-gaming) users struggling with DualStick (Gamepad-like) interfaces, led us to the assumption that participants encounter difficulties when simultaneously and asymmetrically controlling rotation and translation with their hands in space.
Also, as repeatedly observed in the past, users are capable of adapting to ``poorly'' designed interfaces in remarkable ways while still delivering good performance, which then can result, for instance, in a discrepancy between objective and subjective measures (user experience and usability), see \textbf{H3.3}, cf. \cite{Flemming2022}.
None of this is observed in this study.
Our diverse study population in terms of prior experience \changed{and gender (cf. \cite{Peck2020, Peck2021DiVRsifyBT})}, does not exhibit measurable issues with bimanual operation, even the opposite is observable.
The reasons for this can be related to how participants structure the subtasks of translation and rotation in relation to each other. 
This includes whether they are perceived conceptually as a single task and thus executed either sequentially or simultaneously.
Hashemian et al., for instance, conducted a multi-tasking experiment, where participants had to perform a simple object interaction task with one hand while traveling through a virtual environment \cite{Hashemian2023LeaningAndInteraction}. 
Participants performed overall worse in comparison to walking and Leaning interfaces when they had to use their second hand to control travel.
The difference between the observations in both experiments might be that translation and rotation could be mentally fused into a single task more quickly, than object interaction and travel.


\subsection{Orientation Selection vs. Snap Turning}
The central scientific questions of our work revolve around how and whether the positive results from previous studies regarding Orientation Selection \cite{Zielasko2022rotation} can be extrapolated to motion interfaces.
We have already discussed that one of the potentially significant hurdles in the integration with other interfaces, namely, bimanual operation, has not proven to be a problem.
Furthermore, no other issues emerged, and Orientation Selection demonstrated good overall performance across the investigated locomotion paradigms.
Our results allow us to recommend Orientation Selection over Snap Turning, especially when it comes to performance.
However, still more investigation is needed especially when it comes to user experience, presence, and orientation.
Our results on the latter again were not conclusive and we will further discuss this from another perspective in Section~\ref{disc_orientation}.

\subsection{Pointing-directed Steering vs. Leaning}
Previous studies consistently emphasize the clear advantages of Human-Joystick interfaces over controller-based interfaces \cite{Zielasko2016HMDnav, Hashemian2023LeaningAndInteraction, Hashemian2023, Adhikari2021lean}, in the domain of continuous and semi-continuous locomotion. 
However, to the best of our knowledge, there is only one other study~\cite{Hashemian2023LeaningAndInteraction} that compares Human-Joystick interfaces to Pointing-Directed Steering interfaces, which can be argued as one of the simplest and most effective (controller-based) methods of locomotion. 
Thus, we have undertaken this comparison, albeit in an exploratory manner and in the context of discrete rotations. 
We find sufficient evidence to motivate further research on this point and based on our results hypothesize that Pointing-Directed Steering is easier to operate than Leaning interfaces (usability, task load\changed{, and performance}).
This is supported by findings from Griffin et al. \cite{Griffin2018handbusy}.
In a bimanual aiming task combined with hands-free vs. controller-based locomotion, they find higher presence with embodied methods but also a higher cognitive load.
Furthermore, Pointing-Directed Steering is not without embodied/physical cues, which repeatedly come into discussion when, for instance, arguing for potentially enhanced spatial updating capabilities of Leaning interfaces.
In our experiment, we do not find any differences in spatial perception. 
However, we also have a reasonable suspicion that our measuring instrument may not have been powerful enough to capture potential differences. 
We discuss this in the next section.

Hashemian et al. \cite{Hashemian2023LeaningAndInteraction} find partially contradicting results regarding effectiveness and user experience measures, when Pointing-Directed Steering is compared to walking, head-joystick, and a Pivoting interface.
But first, the exact implementation of the Pointing-Directed Steering interface remains unclear, but the use of a thumbstick for speed regulation suggests that it is a more complex implementation than the one presented here.
Second, the difference would not be surprising, considering that they specifically focused on which interfaces work best in conjunction with simultaneous interaction (see Section~\ref{sec_disc_bimanual}), a more specific use case.

Ultimately, body-based motion methods may be the best choice when it comes to fun and simultaneous interaction \cite{Hashemian2023LeaningAndInteraction, Griffin2018handbusy}, while Pointing-Directed Steering is preferred when efficiency and usability take center stage.
This actually makes a lot of sense, as in games, interfaces often involve elements of skill and physical control.
However, as mentioned earlier, at least our observations are based only on semi-continuous interfaces, i.e., with a continuous translation and a discrete rotation. 
Further research is needed for a definitive transfer to continuous interfaces.

\subsection{Task-load and Orientation\label{disc_orientation}}
As we discussed earlier, measuring spatial orientation in navigation tasks is not straightforward.
It often comes down to experiments either being highly constructed and taking place in well-parameterized grid-like environments \cite{Grechkin2014bodybasedOrientation, Sargunam2018, Ragan2017ampHeadRoation} or being based on frequent measurements, which heavily interfere with the experiment \cite{Adhikari2022hyperjump, Zielasko2022rotation}, or they take place in more generalizable environments where effects are hardly observed.
For this reason, we tried to argue that increasing problems with orientation should also be reflected in task load. 
However, we have also discussed that task load is generally influenced by motion interfaces \cite{Lim2022EvaluationOU, Marsh2013CognitiveDO, Marsh2011assessing}, making it challenging to pinpoint exact sources of disruption.
In the following, we want to examine whether our results reveal more.
The descriptive data (see Figure~\ref{img_result_tlx}) and the inferential statistics (see Table~\ref{tab_inference}, Pair 4) show a significantly higher workload for the two Leaning interfaces and otherwise no anomalies.
If the memory-spatialization task stimulated the different interfaces to affect the task difficulty differently, it would suggest that Leaning had a more negative impact than the potential difference between Orientation Selection and Snap Rotation.
If the interfaces did not successfully respond to the stimulus of the memory-spatialization task, it might indicate that using Leaning was maybe just more exhausting, and we can not make any further conclusions about other factors or conditions.
The second interpretation is supported by post-hoc analyses:
No differences in map drawing errors are detectable and the overall NASA TLX score, $r(180) = .072$, $p = .338$, as well as the mental demand subscale, $r(180) = .027$, $p = .715$, show also no indication of a linear correlation with the map drawing error.

\subsection{Limitations}
In addition to the ones already mentioned, our conducted investigations have some limitations, particularly concerning generalizability. 
We will address these separately in the following discussion.

For the sake of better comparability, we make the decision to place the rotation specification in the dominant hand of the users across all interfaces. 
We do not give participants the choice. 
This should not be overlooked, as the dominant hand plays a particularly important role in the human motor model \cite{Guiard1987asymmetric}.

By systematically having placed dead ends that had to be visited in our courses (see Section~\ref{environments}), we provoke the necessity of 180° turns. 
In use cases where such turns are avoidable or not necessary, the performance advantage of Orientation Selection methods will be smaller.

\changed{Because we have anticipated issues with the two-handed use of the interfaces and added 2 more methods to the study design that fall outside our $2 \times 2$ design (translation, rotation), there are overall more Pointing Direct Steering over Leaning methods tested. 
This may have favored the former regarding learning effects (cf. \cite{zielasko2024carryOver}).}

The motivation behind discrete or semi-discrete movement methods primarily revolves around the prevention of cybersickness. However, we do not systematically investigate the potential influence of movement methods on cybersickness in this study, as potential carryover effects would necessitate significant changes in the study design \cite{Zielasko2021sicknessSubject}. 
While it is reasonable to assume that this does not differ between discrete rotation methods, interaction effects in conjunction with continuous translation cannot be entirely ruled out and should be the subject of future research.

\section{Conclusion}
In this study, we delve into the integration of discrete Orientation Selection into continuous steering interfaces, with the dual aim of preserving the seamless perception of real-world movement and mitigating the risk of inducing cybersickness. 
Our research encompasses an exploration of various input mappings, including bimanual interaction as a solution to potential conflicts in standard input mappings. 
We also develop unimanual alternatives, such as employing a Human-Joystick.
The outcomes of our empirical study, primarily focused on a primed search task, yield unexpected and insightful findings:

Our user group, spanning multiple levels of gaming experience, exhibits minimal difficulties with the bimanual, asymmetric interfaces. 
The performance of Orientation Selection is found to be at least on par with the established Snap Rotation, showcasing the feasibility of integrating discrete rotation methods into continuous steering interfaces.

Additionally, our exploratory analysis unearths intriguing insights regarding Pointing-Directed Steering methods. 
Our implementation outperforms embodied interfaces in terms of usability and task load. 
These findings suggest that Pointing-Directed Steering holds the potential as a method that not only provides effective control but also requires a lower level of cognitive effort, potentially enhancing user experiences across various virtual environments.

\bibliographystyle{IEEEtran}
\bibliography{IEEEabrv,main}

\begin{thebibliography}{10}
\providecommand{\url}[1]{#1}
\csname url@samestyle\endcsname
\providecommand{\newblock}{\relax}
\providecommand{\bibinfo}[2]{#2}
\providecommand{\BIBentrySTDinterwordspacing}{\spaceskip=0pt\relax}
\providecommand{\BIBentryALTinterwordstretchfactor}{4}
\providecommand{\BIBentryALTinterwordspacing}{\spaceskip=\fontdimen2\font plus
\BIBentryALTinterwordstretchfactor\fontdimen3\font minus \fontdimen4\font\relax}
\providecommand{\BIBforeignlanguage}[2]{{%
\expandafter\ifx\csname l@#1\endcsname\relax
\typeout{** WARNING: IEEEtran.bst: No hyphenation pattern has been}%
\typeout{** loaded for the language `#1'. Using the pattern for}%
\typeout{** the default language instead.}%
\else
\language=\csname l@#1\endcsname
\fi
#2}}
\providecommand{\BIBdecl}{\relax}
\BIBdecl

\bibitem{Bowman1997travel}
D.~A. Bowman, D.~Koller, and L.~F. Hodges, ``{Travel in Immersive Virtual Environments: An Evaluation of Viewpoint Motion Control Techniques},'' in \emph{Proc. of IEEE VR}, 1997, pp. 45--52.

\bibitem{Riecke2021teleportProCon}
B.~E. Riecke and D.~Zielasko, ``{Continuous vs. Discontinuous (Teleport) Locomotion in VR: How Implications can Provide both Benefits and Disadvantages},'' in \emph{Proc. of IEEE VR Abstracts and Workshops}, 2021, pp. 373--374.

\bibitem{Bakker2003}
N.~H. Bakker, P.~O. Passenier, and P.~J. Werkhoven, ``{Effects of Head-Slaved Navigation and the Use of Teleports on Spatial Orientation in Virtual Environments},'' \emph{Human Factors}, vol.~45, no.~1, pp. 160--169, 2003.

\bibitem{Prithul2021teleportation}
A.~Prithul, I.~B. Adhanom, and E.~Folmer, ``{Teleportation in Virtual Reality; A Mini-Review},'' \emph{Frontiers in Virtual Reality}, p. 138, 2021.

\bibitem{LaViola2000}
J.~J. LaViola, ``{A Discussion of Cybersickness in Virtual Environments},'' \emph{ACM SIGCHI Bull.}, vol.~32, no.~1, p. 47–56, 2000.

\bibitem{Buttussi2023}
F.~Buttussi and L.~Chittaro, ``{Acquisition and Retention of Spatial Knowledge Through Virtual Reality Experiences: Effects of VR Setup and Locomotion Technique},'' \emph{International Journal of Human-Computer Studies}, vol. 177, 2023.

\bibitem{Zielasko2019deskTravel}
D.~Zielasko, B.~Weyers, and T.~W. Kuhlen, ``{A Non-Stationary Office Desk Substitution for Desk-Based and HMD-Projected Virtual Reality},'' in \emph{Proc. of IEEE VR}, 2019, pp. 1884--1889.

\bibitem{Schmelter2020}
T.~Schmelter and K.~Hildebrand, ``{Analysis of Interaction Spaces for VR in Public Transport Systems},'' in \emph{Proc. of IEEE VR Abstracts and Workshops}, 2020, pp. 279--280.

\bibitem{Sargunam2018}
S.~P. Sargunam and E.~D. Ragan, ``{Evaluating Joystick Control for View Rotation in Virtual Reality with Continuous Turning, Discrete Turning, and Field-of-View Reduction},'' in \emph{Proc. of International Workshop on Interactive and Spatial Computing}, 2018, p. 74–79.

\bibitem{Arns2004}
L.~Arns and C.~Cruz-Neira, ``{Effects of Physical and Virtual Rotations and Display Device on Users of an Architectural Walkthrough},'' in \emph{Proc. of ACM SIGGRAPH International Conference on Virtual Reality Continuum and Its Applications in Industry}, 2004, pp. 104--111.

\bibitem{Zielasko2022rotation}
D.~Zielasko, J.~Heib, and B.~Weyers, ``{Systematic Design Space Exploration of Discrete Virtual Rotations in VR},'' in \emph{Prof. of IEEE Virtual Reality and 3D User Interfaces}, 2022, pp. 693--702.

\bibitem{Farmani2018snapping}
Y.~Farmani and R.~J. Teather, ``{Viewpoint Snapping to Reduce Cybersickness in Virtual Reality},'' in \emph{Proc. of Graphics Interface}, 2018, pp. 168 -- 175.

\bibitem{Guy15}
E.~Guy, P.~Punpongsanon, D.~Iwai, K.~Sato, and T.~Boubekeur, ``{LazyNav: 3d Ground Navigation With Non-Critical Body Parts},'' in \emph{Proc. of IEEE Symposium on 3D User Interfaces}, 2015, pp. 43--50.

\bibitem{WangLindemann2012}
J.~Wang and R.~Lindeman, ``{Leaning-Based Travel Interfaces Revisited: Frontal versus Sidewise Stances for Flying in 3D Virtual Spaces},'' in \emph{Proc. of ACM VRST}, 2012, p. 121–128.

\bibitem{ValkovWIM2010}
\BIBentryALTinterwordspacing
D.~Valkov, F.~Steinicke, G.~Bruder, and K.~H. Hinrichs, ``{Traveling in 3D Virtual Environments with Foot Gestures and a Multi-Touch enabled WIM},'' in \emph{Proc. of Virtual Reality International Conference}, 2010, pp. 171--180. [Online]. Available: \url{http://basilic.informatik.uni-hamburg.de/Publications/2010/VSBH10}
\BIBentrySTDinterwordspacing

\bibitem{Marchal2011Joyman}
M.~Marchal, J.~Pettré, and A.~Lécuyer, ``{Joyman: A Human-Scale Joystick for Navigating in Virtual Worlds},'' in \emph{Proc. of IEEE 3DUI}, 2011, pp. 19--26.

\bibitem{Freiberg2017}
J.~Freiberg, A.~Kitson, and B.~E. Riecke, ``{Development and Evaluation of a Hands-Free Motion Cueing Interface for Ground-Based Navigation},'' in \emph{Proc. of IEEE VR}, 2017, pp. 273--274.

\bibitem{Mielhbradt2018}
J.~Miehlbradt, A.~Cherpillod, S.~Mintchev, M.~Coscia, F.~Artoni, D.~Floreano, and S.~Micera, ``{Data-Driven Body–Machine Interface for the Accurate Control of Drones},'' \emph{Proceedings of the National Academy of Sciences}, vol. 115, p. 201718648, 07 2018.

\bibitem{Harris2014}
A.~Harris, K.~Nguyen, P.~T. Wilson, M.~Jackoski, and B.~Williams, ``{Human Joystick: Wii-Leaning to Translate in Large Virtual Environments},'' in \emph{Proc. ACM International Conference on Virtual-Reality Continuum and Its Applications in Industry}, 2014, p. 231–234.

\bibitem{Nguyen19}
T.~Nguyen-Vo, B.~E. Riecke, W.~Stuerzlinger, D.-M. Pham, and E.~Kruijff, ``{NaviBoard and NaviChair: Limited Translation Combined with Full Rotation for Efficient Virtual Locomotion},'' \emph{IEEE Transactions on Visualization and Computer Graphics}, vol.~27, no.~1, pp. 165--177, 2021.

\bibitem{Zielasko2020LookAround}
D.~Zielasko, Y.~C. Law, and B.~Weyers, ``{Take a Look Around – The Impact of Decoupling Gaze and Travel-direction in Seated and Ground-based Virtual Reality Utilizing Torso-directed Steering},'' in \emph{Proc. of IEEE Conference on Virtual Reality and 3D User Interfaces (VR)}, 2020, pp. 398--406.

\bibitem{KitsonRiecke2015}
A.~Kitson, B.~E. Riecke, A.~M. Hashemian, and C.~Neustaedter, ``{NaviChair: Evaluating an Embodied Interface Using a Pointing Task to Navigate Virtual Reality},'' in \emph{Proc. of ACM Symposium on Spatial User Interaction}.\hskip 1em plus 0.5em minus 0.4em\relax Association for Computing Machinery, 2015, p. 123–126.

\bibitem{laViola2017_3DUI}
J.~J. LaViola~Jr, E.~Kruijff, R.~P. McMahan, D.~Bowman, and I.~P. Poupyrev, \emph{{3D User Interfaces: Theory and Practice}}.\hskip 1em plus 0.5em minus 0.4em\relax Addison-Wesley Professional, 2017.

\bibitem{Bozgeyikli2016}
E.~Bozgeyikli, A.~Raij, S.~Katkoori, and R.~Dubey, ``{Point \& Teleport Locomotion Technique for Virtual Reality},'' in \emph{Proc. of ACM Symposium on Computer-Human Interaction in Play}, 2016, p. 205–216.

\bibitem{Adhikari2021lean}
A.~Adhikari, A.~M. Hashemian, T.~Nguyen-Vo, E.~Kruijff, M.~v.~d. Heyde, and B.~E. Riecke, ``{Lean to Fly: Leaning-Based Embodied Flying Can Improve Performance and User Experience in 3D Navigation},'' \emph{Frontiers in Virtual Reality}, vol.~2, 2021.

\bibitem{Bhandari2018dash}
J.~Bhandari, P.~R. MacNeilage, and E.~Folmer, ``{Teleportation without Spatial Disorientation Using Optical Flow Cues},'' in \emph{Graphics Interface}, 2018, pp. 162--167.

\bibitem{Brument2021}
H.~Brument, G.~Bruder, M.~Marchal, A.~H. Olivier, and F.~Argelaguet, ``{Understanding, Modeling and Simulating Unintended Positional Drift during Repetitive Steering Navigation Tasks in Virtual Reality},'' \emph{IEEE TVCG}, vol.~27, no.~11, pp. 4300--4310, 2021.

\bibitem{vonKapri2011}
A.~von Kapri, T.~Rick, and S.~Feiner, ``{Comparing Steering-Based Travel Techniques for Search Tasks in a CAVE},'' in \emph{Proc. of IEEE VR}, 2011, pp. 91--94.

\bibitem{Flemming2022}
C.~Flemming, B.~Weyers, and D.~Zielasko, ``{How to Take a Brake from Embodied Locomotion -- Seamless Status Control Methods for Seated Leaning Interfaces},'' in \emph{Proc. of IEEE Conference on Virtual Reality and 3D User Interfaces (VR)}, 2022.

\bibitem{Adhikari2022hyperjump}
A.~Adhikari, D.~Zielasko, I.~Aguilar, A.~Bretin, E.~Kruijff, M.~von~der Heyde, and B.~E. Riecke, ``{Integrating Continuous and Teleporting VR Locomotion Into a Seamless ‘HyperJump’Paradigm},'' \emph{IEEE TVCG}, 2022.

\bibitem{Zielasko2016HMDnav}
D.~Zielasko, S.~Horn, S.~Freitag, B.~Weyers, and T.~W. Kuhlen, ``{Evaluation of Hands-Free HMD-Based Navigation Techniques for Immersive Data Analysis},'' in \emph{Proc. of IEEE Symposium on 3D User Interfaces}, 2016, pp. 113--119.

\bibitem{Gao2021}
B.~Gao, Z.~Mai, H.~Tu, and H.~B.-L. Duh, ``{Evaluation of Body-centric Locomotion with Different Transfer Functions in Virtual Reality},'' in \emph{In Proc. of IEEE VR}, 2021, pp. 493--500.

\bibitem{Zhixin2016fingerWIP}
Z.~Yan, R.~W. Lindeman, and A.~Dey, ``{Let Your Fingers Do the Walking: A Unified Approach for Efficient Short-, Medium-, and Long-Distance Travel in VR},'' in \emph{Proc. of IEEE 3DUI}, 2016, pp. 27--30.

\bibitem{Nabiyouni2015}
M.~Nabiyouni, A.~Saktheeswaran, D.~A. Bowman, and A.~Karanth, ``{Comparing the Performance of Natural, Semi-natural, and Non-natural Locomotion Techniques in Virtual Reality},'' in \emph{Proc. of 3DUI}, 2015, pp. 3--10.

\bibitem{Ortega2020Gamepad}
F.~R. Ortega, A.~S. Williams, K.~Tarre, A.~Barreto, and N.~Rishe, ``{3D Travel Comparison Study between Multi-Touch and GamePad},'' \emph{International Journal of Human-Computer Interaction}, vol.~36, no.~18, pp. 1699--1713, 2020.

\bibitem{Hashemian2020HeadJoystick}
A.~M. Hashemian, M.~Lotfaliei, A.~Adhikari, E.~Kruijff, and B.~E. Riecke, ``{HeadJoystick: Improving Flying in VR using a Novel Leaning-Based Interface},'' \emph{IEEE TVCG}, 2020.

\bibitem{Suma2007}
E.~A. Suma, S.~Babu, and L.~F. Hodges, ``{Comparison of Travel Techniques in a Complex, Multi-Level 3D Environment},'' in \emph{Proc. of IEEE 3DUI}, 2007.

\bibitem{Mine1995virtual}
M.~R. Mine, ``{Virtual Environment Interaction Technique},'' \emph{UNC Chapel Hill CS Dept}, 1995.

\bibitem{Ruddle2013}
R.~A. Ruddle, E.~Volkova, and H.~H. B\"{u}lthoff, ``{Learning to Walk in Virtual Reality},'' \emph{ACM Transactions on Applied Perception}, vol.~10, no.~2, 2013.

\bibitem{Hashemian2023}
A.~M. Hashemian, A.~Adhikari, E.~Kruijff, M.~v.~d. Heyde, and B.~E. Riecke, ``{Leaning-Based Interfaces Improve Ground-Based VR Locomotion in Reach-the-Target, Follow-the-Path, and Racing Tasks},'' \emph{IEEE TVCG}, vol.~29, no.~3, pp. 1748--1768, 2023.

\bibitem{Jeong2009}
D.~H. Jeong, C.~G. Song, R.~Chang, and L.~Hodges, ``{User Experimentation: An Evaluation of Velocity Control Techniques in Immersive Virtual Environments},'' \emph{Virtual Reality}, vol.~13, pp. 41--50, 2009.

\bibitem{Wilson2016armswing}
P.~T. Wilson, W.~Kalescky, A.~MacLaughlin, and B.~Williams, ``{VR Locomotion: Walking $>$ Walking in Place $>$ Arm Swinging},'' in \emph{Proc. of ACM SIGGRAPH Conference on Virtual-Reality Continuum and Its Applications in Industry - Volume 1}, 2016, p. 243–249.

\bibitem{Freitag2016}
S.~Freitag, B.~Weyers, and T.~W. Kuhlen, ``{Automatic Speed Adjustment for Travel Through Immersive Virtual Environments Based on Viewpoint Quality},'' in \emph{Proc. of IEEE 3DUI}, 2016, pp. 67--70.

\bibitem{Kitson2017}
A.~Kitson, A.~M. Hashemian, E.~R. Stepanova, E.~Kruijff, and B.~E. Riecke, ``{Comparing Leaning-Based Motion Cueing Interfaces for Virtual Reality Locomotion},'' in \emph{Proc. of IEEE Symposium on 3D User Interfaces}, 2017, pp. 73--82.

\bibitem{Hashemian2021IsWN}
A.~M. Hashemian, E.~Kruijff, A.~Adhikari, M.~von~der Heyde, I.~A. Aguilar, and B.~E. Riecke, ``{Is Walking Necessary for Effective Locomotion and Interaction in VR?}'' in \emph{Proc. of IEEE VR Abstracts and Workshops}, 2021, pp. 395--396.

\bibitem{Kruijff16}
E.~Kruijff, A.~Marquardt, C.~Trepkowski, R.~W. Lindeman, A.~Hinkenjann, J.~Maiero, and B.~E. Riecke, ``{On Your Feet! Enhancing Vection in Leaning-Based Interfaces through Multisensory Stimuli},'' in \emph{Proc. of Symposium on Spatial User Interaction}, ser. SUI '16.\hskip 1em plus 0.5em minus 0.4em\relax Association for Computing Machinery, 2016, p. 149–158.

\bibitem{Riecke2006}
B.~E. Riecke, ``{Simple User-Generated Motion Cueing Can Enhance Self-Motion Perception (Vection) in Virtual Reality},'' in \emph{Proc. of ACM Symposium on Virtual Reality Software and Technology}, ser. VRST '06.\hskip 1em plus 0.5em minus 0.4em\relax Association for Computing Machinery, 2006, p. 104–107.

\bibitem{Langbehn2019turn}
E.~Langbehn, J.~Wittig, N.~Katzakis, and F.~Steinicke, ``{Turn Your Head Half Round: VR Rotation Techniques for Situations With Physically Limited Turning Angle},'' in \emph{Proc. of Mensch und Computer}, 2019.

\bibitem{Buttussi2021}
F.~Buttussi and L.~Chittaro, ``{Locomotion in Place in Virtual Reality: A Comparative Evaluation of Joystick, Teleport, and Leaning},'' \emph{IEEE TVCG}, vol.~27, no.~1, pp. 125--136, 2021.

\bibitem{Hashemian2023LeaningAndInteraction}
A.~M. Hashemian, A.~Adhikari, I.~A. Aguilar, E.~Kruijff, M.~v.~d. Heyde, and B.~E. Riecke, ``{Leaning-Based Interfaces Improve Simultaneous Locomotion and Object Interaction in VR Compared to the Handheld Controller},'' \emph{IEEE TVCG}, pp. 1--18, 2023.

\bibitem{Kemeny2017}
A.~Kemeny, P.~George, F.~Merienne, and F.~COLOMBET, ``{New VR Navigation Techniques to Reduce Cybersickness},'' in \emph{{The Engineering Reality of Virtual Reality}}.\hskip 1em plus 0.5em minus 0.4em\relax {Society for Imaging Science and Technology}, 2017, pp. 48--53.

\bibitem{Zielasko2021sicknessSubject}
D.~Zielasko, ``{Subject 001 - A Detailed Self-Report of Virtual Reality Induced Sickness},'' in \emph{Proc. of IEEE VR Abstracts and Workshops}, 2021, pp. 165--168.

\bibitem{fernandes2016}
A.~S. Fernandes and S.~K. Feiner, ``{Combating VR Sickness Through Subtle Dynamic Field-Of-View Modification},'' \emph{Proc. of IEEE 3DUI}, pp. 201--210, 2016.

\bibitem{zielasko2018dynamic}
D.~Zielasko, A.~Mei{\ss}ner, S.~Freitag, B.~Weyers, and T.~W. Kuhlen, ``{Dynamic Field of View Reduction Related to Subjective Sickness Measures in an HMD-based Data Analysis Task},'' in \emph{Proc. of IEEE VR Workshop on Everday Virtual Reality}, 2018.

\bibitem{bolas2017dynamic}
M.~Bolas, J.~A. Jones, I.~Mcdowall, and E.~Suma, ``{Dynamic Field of View Throttling as a Means of Improving User Experience in Head Mounted Virtual Environments},'' 2017, uS Patent 9,645,395.

\bibitem{Benda2023}
B.~Benda, S.~P. Sargunam, M.~Nourani, and E.~D. Ragan, ``{An Evaluation of View Rotation Techniques for Seated Navigation in Virtual Reality},'' \emph{IEEE Transactions on Visualization and Computer Graphics}, pp. 1--14, 2023.

\bibitem{Farmani2020evaluating}
Y.~Farmani and R.~J. Teather, ``{Evaluating Discrete Viewpoint Control to Reduce Cybersickness in Virtual Reality},'' \emph{Virtual Reality}, pp. 1--20, 2020.

\bibitem{Bimberg2021anchor}
P.~Bimberg, T.~Weissker, A.~Kulik, and B.~Froehlich, ``{Virtual Rotations for Maneuvering in Immersive Virtual Environments},'' in \emph{Proceedings of the 27th ACM Symposium on Virtual Reality Software and Technology}, 2021.

\bibitem{Funk2019}
M.~Funk, F.~M\"{u}ller, M.~Fendrich, M.~Shene, M.~Kolvenbach, N.~Dobbertin, S.~G\"{u}nther, and M.~M\"{u}hlh\"{a}user, ``{Assessing the Accuracy of Point \& Teleport Locomotion with Orientation Indication for Virtual Reality Using Curved Trajectories},'' in \emph{Proc. of ACM CHI}, 2019, pp. 1–--12.

\bibitem{Wolf2021AugmentedTeleport}
D.~Wolf, M.~Rietzler, L.~Bottner, and E.~Rukzio, ``{Augmenting Teleportation in Virtual Reality With Discrete Rotation Angles},'' \emph{ArXiv}, vol. abs/2106.04257, 2021.

\bibitem{Griffin2019}
N.~N. Griffin and E.~Folmer, ``{Out-of-Body Locomotion: Vectionless Navigation with a Continuous Avatar Representation},'' in \emph{Proc. of ACM VRST}, 2019.

\bibitem{Lugrin2019}
J.-L. Lugrin, A.~Juchno, P.~Schaper, M.~Landeck, and M.~E. Latoschik, ``{Drone-Steering: A Novel VR Traveling Technique},'' in \emph{Proc. of ACM VRST}, 2019.

\bibitem{feld2024transition}
N.~Feld, P.~Bimberg, B.~Weyers, and D.~Zielasko, ``{Simple and Efficient: Evaluation of Transitions for Task-Driven Cross-Reality Experiences},'' \emph{IEEE Transactions on Visualization and Computer Graphics}, 2024.

\bibitem{woelwer2024posterFadeaway}
M.~Wölwer, B.~Weyers, and D.~Zielasko, ``{How Long Do I Want to Fade Away? The Duration of Fade-To-Black Transitions in Target-Based Discontinuous Travel (Teleportation)},'' \emph{Proc. of IEEE VR Abstracts and Workshops}, 2024.

\bibitem{Swinnen1991control}
S.~Swinnen, D.~Young, C.~Walter, and D.~Serrien, ``{Control of Asymmetrical Bimanual Movements},'' \emph{Experimental Brain Research}, vol.~85, pp. 163--173, 1991.

\bibitem{Balakrishnan2000Symmetric}
R.~Balakrishnan and K.~Hinckley, ``{Symmetric Bimanual Interaction},'' in \emph{Proc. of ACM CHI}, 2000, pp. 33--40.

\bibitem{hart1988development}
S.~G. Hart and L.~E. Staveland, ``{Development of NASA-TLX (Task Load Index): Results of Empirical and Theoretical Research},'' in \emph{Advances in Psychology}.\hskip 1em plus 0.5em minus 0.4em\relax Elsevier, 1988, vol.~52, pp. 139--183.

\bibitem{brooke1996sus}
J.~Brooke, ``{SUS - A Quick and Dirty Usability Scale},'' \emph{Usability Evaluation in Industry}, vol. 189, no. 194, pp. 4--7, 1996.

\bibitem{Keshavarz2011fms}
B.~Keshavarz and H.~Hecht, ``{Validating an Efficient Method to Quantify Motion Sickness},'' \emph{Human Factors}, vol.~53, no.~4, pp. 415--426, 2011.

\bibitem{ShapiroWilk65}
\BIBentryALTinterwordspacing
S.~S. Shapiro and M.~B. Wilk, ``{An Analysis of Variance Test for Normality (Complete Samples)},'' \emph{Biometrika}, vol.~52, no. 3/4, pp. 591--611, 1965. [Online]. Available: \url{http://www.jstor.org/stable/2333709}
\BIBentrySTDinterwordspacing

\bibitem{fisher1956statistical}
R.~A. Fisher, \emph{{Statistical Methods and Scientific Inference}}.\hskip 1em plus 0.5em minus 0.4em\relax Edinburgh, Oliver and Boyd, 1956.

\bibitem{cohen2013statistical}
J.~Cohen, \emph{{Statistical Power Analysis for the Behavioral Sciences}}.\hskip 1em plus 0.5em minus 0.4em\relax Academic press, 2013.

\bibitem{Bangor2009}
A.~Bangor, P.~Kortum, and J.~Miller, ``{Determining What Individual SUS Scores Mean: Adding an Adjective Rating Scale},'' \emph{Journal of Usability Studies}, vol.~4, no.~3, p. 114–123, 2009.

\bibitem{Coluccia2004}
E.~Coluccia and G.~Louse, ``{Gender Differences in Spatial Orientation: A Review},'' \emph{Journal of Environmental Psychology}, vol.~24, no.~3, pp. 329--340, 2004.

\bibitem{Munion2019gender}
A.~K. Munion, J.~K. Stefanucci, E.~Rovira, P.~Squire, and M.~Hendricks, ``{Gender Differences in Spatial Navigation: Characterizing Wayfinding Behaviors},'' \emph{Psychonomic Bulletin \& Review}, vol.~26, pp. 1933--1940, 2019.

\bibitem{Talvas2014bimanualSurvey}
A.~Talvas, M.~Marchal, and A.~Lecuyer, ``{A Survey on Bimanual Haptic Interaction},'' \emph{IEEE Transactions on Haptics}, vol.~7, no.~3, pp. 285--300, 2014.

\bibitem{Owen2005gets}
R.~Owen, G.~Kurtenbach, G.~Fitzmaurice, T.~Baudel, and B.~Buxton, ``{When it gets more difficult, use both hands: exploring bimanual curve manipulation},'' in \emph{Proc. of Graphics Interface}, 2005, pp. 17--24.

\bibitem{obhi2004bimanual}
S.~S. Obhi, ``{Bimanual Coordination: An Unbalanced Field of Research},'' \emph{Motor Control}, vol.~8, no.~2, pp. 111--120, 2004.

\bibitem{Peck2020}
T.~C. Peck, L.~E. Sockol, and S.~M. Hancock, ``{Mind the Gap: The Underrepresentation of Female Participants and Authors in Virtual Reality Research},'' \emph{IEEE Transactions on Visualization and Computer Graphics}, vol.~26, no.~5, pp. 1945--1954, 2020.

\bibitem{Peck2021DiVRsifyBT}
T.~C. Peck, K.~A. McMullen, and J.~Quarles, ``{DiVRsify: Break the Cycle and Develop VR for Everyone},'' \emph{IEEE Computer Graphics and Applications}, vol.~41, pp. 133--142, 2021.

\bibitem{Griffin2018handbusy}
N.~N. Griffin, J.~Liu, and E.~Folmer, ``{Evaluation of Handsbusy vs Handsfree Virtual Locomotion},'' in \emph{Proc. of ACM CHI PLAY}, 2018, pp. 211--–219.

\bibitem{Grechkin2014bodybasedOrientation}
T.~Y. Grechkin and B.~E. Riecke, ``{Re-Evaluating Benefits of Body-Based Rotational Cues for Maintaining Orientation in Virtual Environments: Men Benefit From Real Rotations, Women Don’t},'' in \emph{Proc. of ACM Symposium on Applied Perception}, 2014, pp. 99--102.

\bibitem{Ragan2017ampHeadRoation}
E.~D. Ragan, S.~Scerbo, F.~Bacim, and D.~A. Bowman, ``{Amplified Head Rotation in Virtual Reality and the Effects on 3D Search, Training Transfer, and Spatial Orientation},'' \emph{IEEE TVCG}, vol.~23, no.~8, pp. 1880--1895, 2017.

\bibitem{Lim2022EvaluationOU}
D.~Lim, S.~Shirai, J.~Orlosky, P.~Ratsamee, Y.~Uranishi, and H.~Takemura, ``Evaluation of user interfaces for three-dimensional locomotion in virtual reality,'' \emph{Proceedings ACM SUI}, 2022.

\bibitem{Marsh2013CognitiveDO}
W.~E. Marsh, J.~W. Kelly, V.~J. Dark, and J.~H. Oliver, ``{Cognitive Demands of Semi-Natural Virtual Locomotion},'' \emph{PRESENCE: Teleoperators and Virtual Environments}, vol.~22, pp. 216--234, 2013.

\bibitem{Marsh2011assessing}
------, ``{Assessing the Use of Cognitive Resources in Virtual Reality},'' in \emph{HCI International 2011--Posters’ Extended Abstracts}.\hskip 1em plus 0.5em minus 0.4em\relax Springer, 2011, pp. 120--124.

\bibitem{Guiard1987asymmetric}
Y.~Guiard, ``{Asymmetric Division of Labor in Human Skilled Bimanual Action: The Kinematic Chain as a Model},'' \emph{Journal of Motor Behavior}, vol.~19, no.~4, pp. 486--517, 1987.

\bibitem{zielasko2024carryOver}
D.~Zielasko, B.~Rehling, D.~Clement, and G.~Domes, ``{Carry-Over Effects Ruin Your (Cybersickness) Experiments and Balancing Conditions Is Not a Solution},'' \emph{Proc. of IEEE VR Abstracts and Workshops}, pp. 1--5, 2024.

\end{thebibliography}


\end{document}